\documentclass{svproc}
\usepackage[ruled]{algorithm2e} 

\usepackage[tight]{subfigure}
\usepackage[T1]{fontenc}
\usepackage{textcomp}
\usepackage{graphicx} 
\usepackage{pgfplots}
\usepackage{caption}
\usepackage{graphicx}
\usepackage{bbm}
\usepackage{threeparttable}
\usepackage{soul}
\usepackage[misc]{ifsym}
\usepackage[tight]{subfigure}
\usepackage[hidelinks]{hyperref}

\usepackage{xurl}

\begin{document}
\mainmatter              
\title{Cryptocurrency Valuation: An Explainable AI Approach\thanks{The authors are by the alphabetical order of the last names and both authors are joint first and corresponding authors. \
\textbf{Acknowledgments}: We thank Prof. Campbell Harvey, Prof. Lin William Cong, and Prof. Ye Li for their insightful comments. Luyao thank \href{http://www.fields.utoronto.ca/activities/21-22/mathematical-finance}{Corporate Finance Institute (CFI) Workshop on Mathematical Finance and Cryptocurrencies at the University of Toronto's Fields Institute}, the \href{https://www.glofin.org/}{29th Annual Global Finance Conference}, and \href{https://sites.google.com/view/2022-workshop-explainable-ai}{Workshop on Explainable AI in Finance} and \href{https://sites.google.com/view/women-in-ai-finance}{Workshop on Women in AI and Finance} co-located with the 3rd ACM International Conference on AI in Finance (ICAIF) for hosting her presentations and inspiring discussions among the participants. We thank the anonymous referees at Computing Conference for their professional and thoughtful comments. The corresponding author Luyao Zhang is supported by National Science Foundation China on the project entitled “Trust Mechanism Design on Blockchain: An Interdisciplinary Approach of Game Theory, Reinforcement Learning, and Human-AI Interactions.” (Grant No. 12201266). Luyao Zhang is also with SciEcon CIC, a not-for-profit organization aiming at cultivating interdisciplinary research of both profound insights and practical impacts in the United Kingdom. Yulin Liu is also with Shiku Foundation and Bochsler Finance, Switzerland. We thank the anonymous referees at Computing Conference for their professional and thoughtful comments. }}
\titlerunning{Cryptocurrency Valuation: An Explainable AI Approach}  
%
\author{Yulin Liu\inst{*}\inst{2} \and Luyao Zhang\inst{*}\inst{1}}
\authorrunning{Y. Liu and L. Zhang} 

\institute{Data Science Research Center and Social Science Division, Duke Kunshan University, Suzhou, Jiangsu, 215316, China,\\
\email{corresponding email: lz183@duke.edu}
\and
SciEcon CIC, London, United Kingdom WC2H 9JQ\\
\email{corresponding email: yulin.liu@sciecon.org}}

\maketitle       
\begin{abstract}
Currently, there are no convincing proxies for the fundamentals of cryptocurrency assets. We propose a new market-to-fundamental ratio, the price-to-utility (PU) ratio, utilizing unique blockchain accounting methods. We then proxy various existing fundamental-to-market ratios by Bitcoin historical data and find they have little predictive power for short-term bitcoin returns. However, PU ratio effectively predicts long-term bitcoin returns than alternative methods. Furthermore, we verify the explainability of PU ratio using machine learning. Finally, we present an automated trading strategy advised by the PU ratio that outperforms the conventional buy-and-hold and market-timing strategies. Our research contributes to explainable AI in finance from three facets: First, our market-to-fundamental ratio is based on classic monetary theory and the unique UTXO model of Bitcoin accounting rather than ad hoc; Second, the empirical evidence testifies the buy-low and sell-high implications of the ratio; Finally,  we distribute the trading algorithms as open-source software via Python Package Index for future research, which is exceptional in finance research.
\keywords{Asset Valuation, Machine Learning, Bitcoin, Cryptocurrency, Market Timing, Automated Trading, Explainable AI, Store of Value, UTXO, Python, Open Source}
\end{abstract}

\section{Introduction}
The market cap of cryptocurrency or crypto tokens \footnote{In this article, these two terms are synonyms. The term “token” has a few different meanings in general use, which are beyond the scope of this paper. See Cong and Xiao (2021)~\cite{cong2021categories} for token categories.}  has steadily increased from nearly zero to more than \$1 trillion\footnote{Data source: \href{https://coinmarketcap.com}{https://coinmarketcap.com}}  in the last decade \cite{hardle2020understanding,haeringer2018bitcoin,halaburda2022microeconomics,harvey2021defi}. In addition, institutional and private investors are adding cryptocurrency to their portfolios, and retailers are increasingly accepting cryptocurrency for payment. As most cryptocurrencies have very volatile prices, it would be intriguing to investigate the fundamentals of a cryptocurrency to determine an optimal indicator that reflects its market valuation relative to its fundamentals—that is, is the cryptocurrency overvalued or undervalued? Currently, the existing literature is far from identifying a reliable proxy for token fundamentals.

Drawing on classic monetary theory, we first propose a new market-to-fundamental ratio, which we call the price-to-utility (PU) ratio, utilizing factors specific to cryptocurrency markets. As cryptocurrency is one type of digital currency, we refer to canonical monetary economics as a theoretical foundation to pin down the fundamentals of cryptocurrencies. More specifically, we construct fundamental proxies considering three major currency functions: medium-of-exchange, store-of-value, and unit-of-account for cryptocurrencies. Innovatively, we proxy store-of-value by staking ratios, available for the sake of the unique unspent transaction outputs (UTXO) accounting on the Bitcoin Blockchain \cite{liu2021deciphering}. Then, we construct proxies for a variety of fundamental-to-market ratios using bitcoin (BTC) historical data. We find that different from the stock market, fundamental-to-market ratios have little predictive power on short-term BTC returns, consistent with existing literature. However, PU ratio, together with several other ratios approximating user adoption to fundamentals \cite{cong2021tokenomics}, effectively predict long-term BTC returns than alternative methods.
\par
An effective valuation ratio can thus inform investment strategies: to buy (sell) when the asset is undervalued (overvalued). Theoretically, when the market-to-fundamentals ratio is high (low), the asset is overvalued (undervalued). We use machine learning to verify that the PU ratio has explainability for future returns. We found that buying low and selling high based on PU ratio valuation are associated with higher investment returns than other strategies. 
Finally, we propose an automated trading strategy guided by PU ratio and find that it outperforms conventional buy-and-hold and market-timing strategies.
\par
Our research contributes to explainable AI in finance from three facets: first, our market-to-fundamental ratio is based on classic monetary theory and the unique UTXO model of Bitcoin accounting rather than ad hoc; second, the empirical evidence testifies the buy-low and sell-high implications of the ratio; finally,  we distribute the trading algorithms as open-source software via Python Package Index for future research, which is exceptional in finance research.
\par
We organize the rest of the paper as follows. Subsections 1.1 and 1.2 introduce related literature and the data basics respectively. Section 2 introduces the methods of cryptocurrency valuations and test compares the performance of alternative methods in predictions. Section 3 further evaluates the explainability of the PU-ratio using machine learning methods. Section 4 demonstrates and compares the automated trading strategies. Section 5 concludes and discusses.
\begin{figure}[!htbp]
    \centering
    \includegraphics[width=\linewidth]{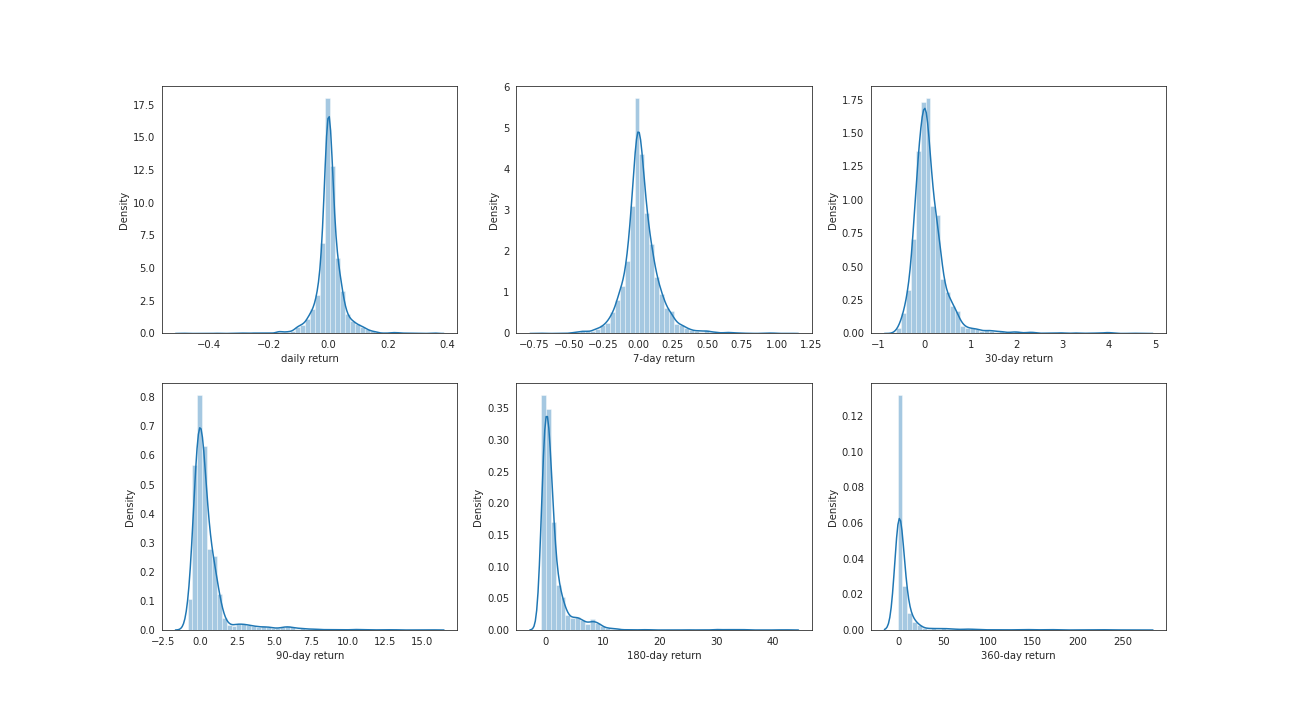}
    \caption{BTC Return Distributions}
    \caption*{Notes: This figure plots the distributions of daily, weekly, 30-day, 90-day, 180-day, and 360-day BTC returns from July 13, 2011, to December 31, 2020 ($3,460$ days in total).}
    \label{fig2}
\end{figure}
\subsection{Related Literature}
Our research contributes to three lines of literature: cryptocurrency valuation, machine learning for economics, and explainable AI in finance.  
\subsubsection{Cryptocurrency Valuation}
Our research adds to the literature on cryptocurrency valuations. Some of the literature provides partial evidence for the validity of our indicators by relating BTC prices to a proxy for one of the monetary functions. For instance, Biais et al. (2018)~\cite{biais2020equilibrium} provide an overlapping generation equilibrium model and emphasize BTC’s role as a medium of exchange. By using a dynamic asset pricing model, Cong, Li, and Wang (2021)~\cite{cong2021tokenomics} derive the value of platform tokens by aggregating heterogeneous users’ demand for cryptocurrency as a medium of exchange. Alabi (2017, 2020)~\cite{alabi2017digital,alabi2020} and Wheatley et al. (2019)~\cite{wheatley2019bitcoin} use Metcalfe’s law~\cite{metcalfe2013metcalfe} to measure the fundamental value of Bitcoin. Metcalfe’s law stipulates that the value of a network is proportional to the square of the number of its active users. While this law captures the value of a given token as a medium of exchange by measuring the number of its active users, it ignores the relatively inactive users who hold cryptocurrencies as a store of value. Athey et al. (2016)~\cite{athey2016bitcoin} document how trading volume, another proxy for the medium of exchange, affects the token price. Their results also provide evidence that BTC is mostly used as a store of value, which underscores the necessity of considering stores of value in BTC valuations. Additionally, current literature has found other predictors of BTC prices, such as attractiveness and popularity on social media,\footnote{In exploring the impact of social media on bitcoin price, Mai et al. (2018)~\cite{mai2018does} and Georgoula et al. (2015)~\cite{georgoula2015using} find that bullish forum sentiment is positively correlated with the price of bitcoin futures, while Polasik et al. (2015)~\cite{polasik2015price} find in an empirical study of Bitcoin’s payment and investment functions that bitcoin price is mainly driven by its popularity on social media. Ciaian, Rajcaniova, and Kancs (2015)~\cite{ciaian2016economics} find that the main drivers of the bitcoin price are the market forces of the bitcoin supply and demand and the attractiveness of Bitcoin for investors and users. Liu and Tsyvinski (2021)~\cite{liu2021deciphering} show that high investor attention, as indicated by factors such as increased Google searches, predicts high future returns over 1- to 2-week horizons for BTC. Borri (2019)~\cite{borri2019conditional} finds that cryptocurrencies are highly correlated with each other, but poorly correlated with other global assets, including gold. Blau (2017)~\cite{blau2017price} identifies that speculative trading does not contribute to the unprecedented rise and subsequent crash in BTC prices.} which are not directly related to the fundamentals considered in our research.\footnote{Fantazzini et al. (2016)~\cite{fantazzini2016everything}, Fry and Cheah (2016)~\cite{fry2016negative}, Corbet, Lucey, and Yarovaya (2018)~\cite{corbet2018datestamping}, Fry (2018)~\cite{fry2018booms}, and references therein used methods developed for stock markets to test bubbles in the cryptocurrency market. The fundamentals used are a function of historical prices, which differs from our monetary theory approach.} 
\subsubsection{Machine Learning for Economics}
Our research also contributes to the emerging literature that applies machine learning to economics in a variety of fields to improve prediction and inform decisions (e.g., Athey 2017, 2019~\cite{athey2017beyond,athey201921}; Athey and Imbens 2017, 2019~\cite{athey2017state,athey2019machine}; Mullainathan and Spiess 2017~\cite{mullainathan2017machine}). Machine learning techniques are shown to constitute a powerful analytical tool for big data and are effective in modeling complex relationships (see Varian 2014~\cite{varian2014big}). Among these techniques, the most relevant to our research are those that can be applied to return predictions in the stock market. For instance, Gu, Kelly, and Xiu (2020)~\cite{gu2020empirical} report evidence on significant returns to investments using machine learning predictions that in some cases offer twice the performance of leading regression-based strategies. Gu, Kelly, and Xiu (2021)~\cite{gu2021autoencoder} developed an autoencoder asset pricing model that first applies advanced unsupervised machine learning methods in finance and delivers much lower out-of-sample pricing errors than traditional factor models. Our research differs from the preceding studies by applying machine learning to the valuation of cryptocurrency; that is, the previous research uses machine learning to advance asset pricing in the stock markets, while ours uses machine learning to pioneer asset pricing in the cryptocurrency market. Moreover, existing literature in general uses supervised machine learning methods in asset pricing, while ours shows how unsupervised machine learning can also inform economics.
\subsubsection{Explainable AI in Finance}
Furthermore, by integrating economic theory into machine learning, our research joins the emerging literature in explainable artificial intelligence (XAI) (e.g., Gunning 2019~\cite{gunning2019xai}; Adadi and Berrada 2018)~\cite{adadi2018peeking}.\footnote{Cong et al. (2020,2021)~\cite{cong2020alphaportfolio,cong2021deep} propose an explainable AI approach for portfolio management in finance.}  XAI critiqued the application of machine learning to predictions and decision-making and described it as a black box that is not understandable by human experts and not perceived as trustworthy by stakeholders. Our study provides three solutions. First, we check the consistency in the economic intuition and algorithm results. For instance, we verify that the function of our simple indicator to inform a “buy-low and sell-high” strategy is, in theory, aligned with the unsupervised machine learning analysis of the BTC data. Second, using one supervised machine learning method with high explainability, we show that the simple indicator alone is efficient in predicting bull markets. Third, we propose an efficient automated trading strategy with high interpretability based on the simple indicator alone for further applications of automation in the industry. In spirits of Explainable AI in Finance, our study also contributes to the literature in finance in comparing the passive investment strategy of buy-and-hold\footnote{Buy-and-hold is an investment strategy in which the investor may use an active strategy to select securities or funds but then lock them in to hold them for a long term.}  and the active investment strategy of market-timing\footnote{Market-timing is an investment strategy in which a market participant attempts to beat the market by predicting its movements and buying and selling accordingly.} (e.g., Malkiel 2003~\cite{malkiel2003passive}; Shilling 1992~\cite{shilling1992market}; Sharpe 1975~\cite{sharpe1975likely}). Existing literature shows that a long-term buy-and-hold strategy tends to outperform the market-timing strategy. Since much of the market’s greatest returns or declines are concentrated in a short time frame, the market-timing strategy often makes use of high-frequency trading technology with tens of thousands of trades per second. However, trading directly on the Bitcoin network is very slow. Depending on the network congestion, the finality time usually lasts one hour or longer. Trading on exchanges involves high trading fees (e.g., the largest crypto exchange Binance charges 0.1\% trading fee) and third-party risk (e.g., deposit hacking, exchange bankruptcy, and internal manipulation by the insiders). Our automated trading strategy involves as few trades as possible, while achieving higher returns than the conventional buy-and-hold strategy. Moreover, other underlying disadvantages of market-timing strategies such as maintenance, labor cost, and the opportunity cost of daily attention are also less of a concern in our trading strategy. To elaborate further, since our approach is an automated trading strategy, it can be delegated to decentralized exchanges (DEXs).\footnote{DEX is a smart contract. Smart contracts are computer codes that can be automatically executed when certain conditions are fulfilled. A smart contract does not need a centralized entity to maintain its functions. Once deployed, it can run timely and objectively, as long as the underlying platform, such as Ethereum, is secure and decentralized.}  A DEX is publicly verifiable and transparent computer code that executes orders automatically when conditions are satisfied. Therefore, users could connect their wallets to a DEX and set their trading strategies in advance.
\subsection{The Data and Basic Characteristics}
Bitcoin is the first cryptocurrency\footnote{Bitcoin was launched on January 3rd, 2009.} and has the largest market share.\footnote{Since its birth, Bitcoin’s market share is mostly more than half of the total cryptocurrency market cap. See \href{https://coinmarketcap.com}{https://coinmarketcap.com}.}  Thus, in this article, we construct and evaluate valuation proxies using bitcoin historical data. The data are sampled daily and downloaded from three open source platforms: CoinMetrics (CM),\footnote{\href{https://coinmetrics.io/data-downloads-2}{https://coinmetrics.io/data-downloads-2}}  the Coin MarketCap (CMC),\footnote{\href{https://coinmarketcap.com/currencies/bitcoin/historical-data}{https://coinmarketcap.com/currencies/bitcoin/historical-data}}  and Google Bigquery.\footnote{\href{https://www.kaggle.com/bigquery/bitcoin-blockchain}{https://www.kaggle.com/bigquery/bitcoin-blockchain}}  We include the historical data on CM from January 3, 2009, to December 31, 2020, and on CMC from April 29, 2013, to December 31, 2020, since April 29, 2013, is the first date that BTC has the off-chain transactions recorded by CMC. The data from Bigquery are from January 3, 2009, to August 31, 2020. The total volume we used in this research is the sum of the on-chain exchange volume collected from CM and the off-chain exchange volume from CMC. 
\par
We now document the main statistical properties of the time series for the BTC returns. Figure \ref{fig2} shows the return distributions of BTC at daily, weekly, 30-day, 90-day, 180-day, and 360-day frequencies. Table \ref{table1} shows the statistics of the BTC returns comparable to Table 1 in Liu and Tsyvinski (2021)~\cite{liu2021risks}. Panel A shows the mean, standard deviation, Sharpe ratio, and kurtosis of BTC return at various frequencies and the percentage of positive returns. The skewness is positive at all frequencies except for daily returns. Panel B shows that the BTC market is highly volatile. The probability of a $5\%$ drop in a day is more than $7\%$ and $2\%$ for a $10\%$ daily loss. However, the probability of a positive movement is larger than a negative movement of the same level.~\footnote{Readers can refer to the working paper version on arXiv for an Appendix: \href{https://arxiv.org/abs/2201.12893}{https://arxiv.org/abs/2201.12893}. The data and code are available on GitHub: \href{https://github.com/SciEcon/CV_XAI}{https://github.com/SciEcon/CV\_XAI}.}

\begin{table}[!htbp] 
\centering
  \caption{Summary Statistics}
  \caption*{Panel A: Summary statistics by various frequencies}
  \label{table1}
  \begin{tabular}{|c|c|c|c|c|c|c|c|}
  \hline
    & Mean & SD & t-Stat & Sharpe & Skewness & Kurtosis & \%>0\\
    \hline
daily & 0.33\% & 4.71\% & 4.21 & 0.07 & -0.21\% & 13.33 & 49.23 \\
return & & & & & & & \\
\hline
weekly & 2.36\% & 13.20\% & 7.72 & 0.18 & 1.12\% & 7.29 & 51.46 \\
return & & & & & & & \\
\hline
30-days & 11.83\% & 43.06\% & 11.50 & 0.27 & 4.16\% & 28.81 & 52.35 \\
return & & & & & & & \\
\hline
90-days & 51.75\% & 139.30\% &15.55 & 0.37 & 4.19\% & 23.87 & 56.54 \\
return & & & & & & & \\
\hline
180-days & 139.43\% & 326.14\% & 17.88 & 0.43 & 5.83\% & 50.87 & 60.61 \\
return & & & & & & & \\
\hline
360 days & 808.55\% & 2485.32\% & 13.57 & 0.33 & 5.99\% & 41.54 & 71.49 \\
return & & & & & & & \\
\hline
\end{tabular}
\end{table}

\begin{table}[!htbp] 
\centering
 \caption*{Panel B. Extreme events of daily returns}
    \begin{tabular}{ccc|ccc}
     \hline
     Disasters & Counts & \% & Miracles & Counts & \% \\
     \hline
     <-5\% & 255 & 7.37 & >5\% & 309 & 8.93\% 
     \\
     <-10\% & 70 & 2.02 & >10\% & 96 & 2.77\% 
     \\
     <-20\% & 12 & 0.35 & >20\% & 16 & 0.46\%
     \\
     <-30\% & 4 & 0.12 & >30\% & 4 & 0.12\%
     \\
     \hline
    \end{tabular}
\caption*{Notes: This table documents the summary statistics of the BTC returns. Panel A reports the daily, weekly, and $30$-day, $90$-day, $180$-day, and $360$-day summary statistics of BTC returns. The mean, standard deviation, $t$-statistics (Newey-West adjusted with $n-1$ lags), Sharpe ratio, skewness, kurtosis, and the percentage of observations that are positive are reported. Panel B reports the percentage of extreme events based on the daily BTC returns. The Bitcoin returns are from July 13, 2011, to December 31, 2020 ($3,460$ days in total).}
\end{table}
\section{Cryptocurrency Valuation Ratios}
In Section 2.1 we present the existing cryptocurrency valuation ratios in both academic literature and industry practice. Section 2.2 introduces the PU ratio and its implications. Section 2.3 assesses the predictive power of various fundamental-to-market ratios on future BTC returns.
\subsection{Industry Cryptocurrency Valuation Ratios}
An optimal valuation ratio should reflect the position of the market valuation of the cryptocurrency relative to its fundamentals. When the fundamental-to-market value is high/low, then the asset is undervalued/overvalued. The market valuation of a cryptocurrency can be precisely measured by its market cap. However, the fundamentals are difficult to determine. Liu and Tsyvinski (2021)~\cite{liu2021risks} construct proxies for the fundamental-to-market ratio in the cryptocurrency market. Among these proxies, the (negative) past 100-week cumulative return (Past100) is based on a strong correlation between the fundamental-to-market value and the negative of the cumulative past returns \cite{de1987further,FAMA1993,moskowitz2021asset}. The other set of measures, including the user-to-market ratio (UMR), the address-to-market ratio (AMR), the transaction-to-market ratio (TMR), and the payment-to-market ratio (PMR), are inspired by the dynamic cryptocurrency asset pricing model developed in Cong, Li, and Wang (2021)~\cite{cong2021tokenomics} and proxy fundamentals by user adoptions. Below, we review other cryptocurrency valuation ratios adopted in industry practice including price-to-earnings (PE) ratio, network-value-to-transactions (NVT) ratio, and price to Metcalfe’s (PM) ratio, and discuss their pros and cons.\footnote{Note that those are market-to-fundamental ratios and the inverse of those are fundamental-to-market ratios.} 
\subsubsection{Price-to-Earning (PE) ratio}
Earnings per share is a widely used proxy for the fundamentals of a company’s stock by financial practitioners. Dividing the stock price by its earnings per share yields the well-known PE ratio (Shiller 2000)~\cite{shiller2000irrational}. Usually, a high PE ratio indicates overvaluation of the stock, and a low ratio indicates undervaluation. Similarly, one could use miners’ earnings as the proxy of the token fundamentals. Miners’ earnings consist of the block rewards and transaction fees paid by users. However, using transaction fees and block rewards to proxy for fundamentals is misleading because these fees are earned by block makers (i.e., miners) and not by token holders. Furthermore, using transaction fees could lead to ludicrous results. For example, a low P/E ratio, resulting from a hike in transaction fees, implies an undervaluation of the token. However, in contrast, it is intuitive that high transaction fees usually impede a more universal adoption of cryptocurrencies and indicate an overvaluation of tokens instead.

\begin{figure}[h]
    \centering
    \includegraphics[width=\linewidth]{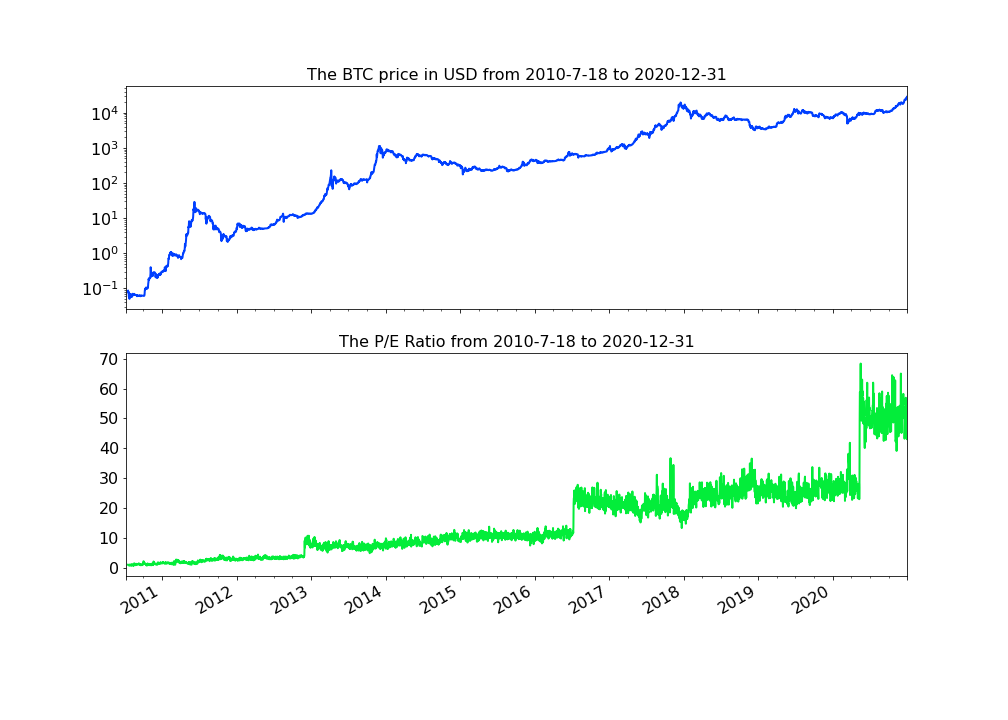}
    \caption{The BTC P/E Ratio and Price in USD}
    \label{fig4}
\end{figure}

Figure \ref{fig4} shows that the BTC PE ratio does not inform either BTC valuation or price movement. The BTC price moves up and down, while the PE ratio increases quasi-linearly with three abrupt jumps. The PE ratio is not a good indicator for assessing BTC valuation.\footnote{BTC supply increases steadily while the block rewards are halved every four years, which explains the quasi-linear growth of the BTC P/E ratio and the three jumps on the halving dates.}
\subsubsection{Network-Value-to-Transactions (NVT) ratio}
\begin{figure}[h]
    \centering
    \includegraphics[width=\linewidth]{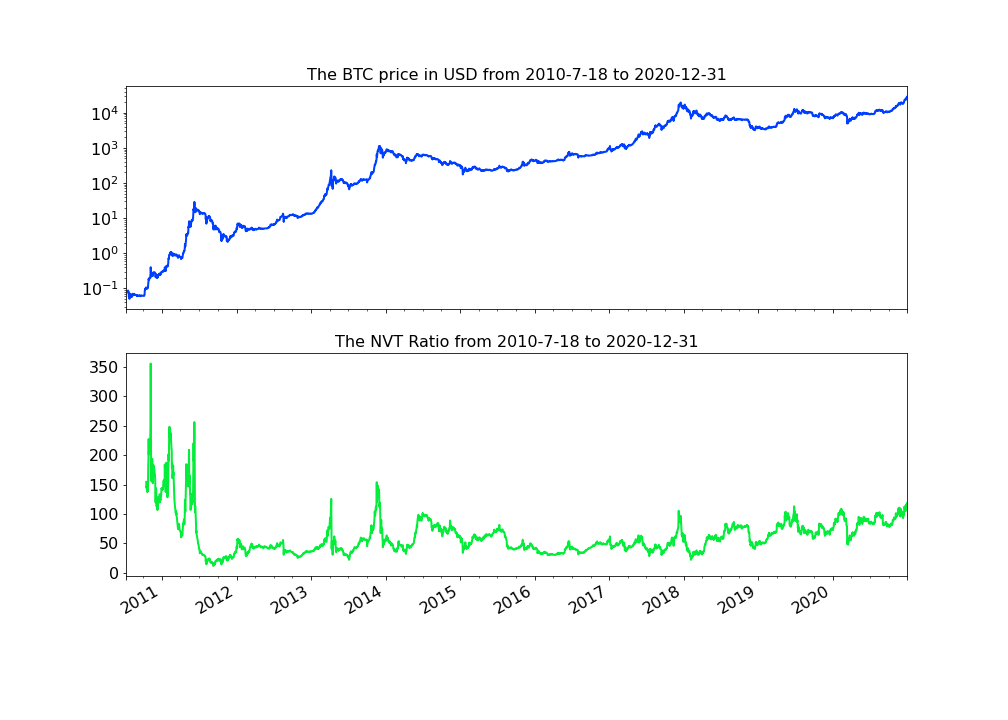}
    \caption{The BTC Price and NVT Ratio in USD}
    \label{fig5}
\end{figure}

NVT ratio was introduced by Willy Woo, an industry pioneer of on-chain analysis.\footnote{\href{https://academy.glassnode.com/indicators/nvt/nvt-ratio}{https://academy.glassnode.com/indicators/nvt/nvt-ratio}} It is defined as the ratio of the market value to token transaction volume in USD over the past 24 hours. That is, the fundamental value of a token derives from how frequently it is used for transactions (i.e., from its function as a medium of exchange). A high NVT ratio indicates that the market cap of the token outpaces the transacted value on its blockchain, indicating speculation. A low NVT ratio caused by an increasing transaction volume indicates a relatively high usage of the token, which, in turn, serves as a signal to buy the undervalued token. The limitation of the NVT ratio lies in the assumption that the fundamental value of a cryptocurrency derives only from its function as a medium of exchange. The usage of cryptocurrencies as a store of value is completely overlooked in this model. To illustrate the point, an increase of the NVT ratio due to a reduction of the transaction volume does not necessarily imply an overvaluation of the token. It could also be the result of an increasing number of long-term holders hoarding more tokens, thus causing a decline in transaction volume. In other words, the NVT ratio offers a glimpse of what is happening, but it is not completely informative regarding buying and selling decisions. This explains why the NVT ratio only captures short-term noises and is not indicative of the general pattern of BTC price movement (Figure \ref{fig5}).

\begin{figure}[h]
    \centering
    \includegraphics[width=\linewidth]{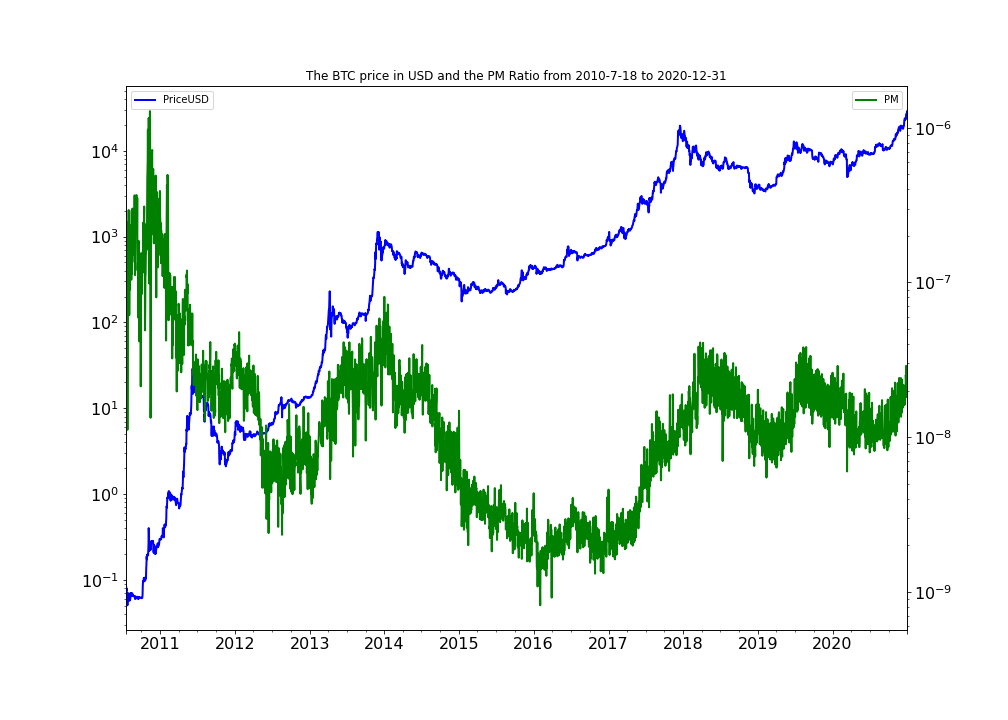}
    \caption{BTC Price in USD and PM Ratio}
    \label{fig6}
\end{figure}
\subsubsection{Price to Metcalfe’s (PM) ratio:}
In the 1980s, Robert Metcalfe, the inventor of Ethernet, proposed that the value of a network is proportional to the square of the number of its users \cite{shapiro1998information}. The underlying logic of this proposal is that the number of connections of a network with n nodes is $\frac{n(n-1)}{2}$, which is asymptotically proportional to $n^2$. Since its inception, Metcalfe’s law has become an influential tool for studying network effects. Figure \ref{fig6} represents the PM ratio counting the number of active addresses as n. Compared with PE ratio and NVT ratio, PM ratio has better potential as a valuation indicator for cryptocurrencies. However, it treats users with different amounts of tokens indifferently and overlooks the inactive users who hoard cryptocurrencies as a store of value.
\subsection{Price-to-Utility Ratio}
A comprehensive assessment of the fundamentals of a cryptocurrency should incorporate all the utilities it provides as a currency, namely, medium of exchange, store of value,\footnote{In this article, “store of value” refers to the function of a token to keep or increase its purchasing power over time.}  and unit of account. In this section, we propose the price-to-utility (PU) ratio that consists of proxies for all three utilities. We define Token Utility (TU) and their proxies as in Equation~\ref{eq:1}:
\begin{equation}
\label{eq:1}
    Token \ utility = \frac{token\ velocity \times staking\ ratio}{price \ volatility \times dilute\  rate} \label{1}
\end{equation}

\begin{figure}[h]
    \centering
    \includegraphics[width=\linewidth]{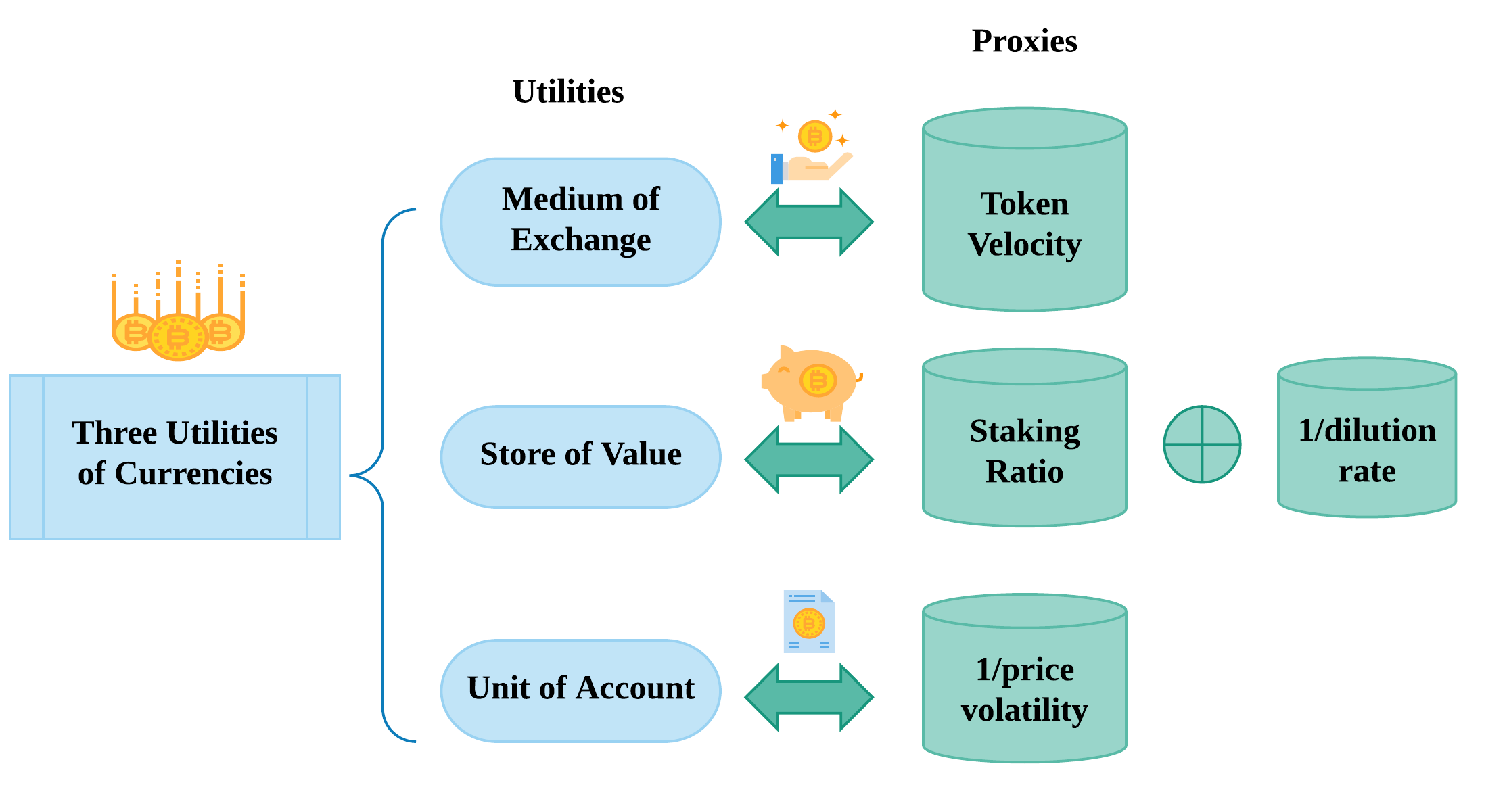}
    \caption{The Three Utilities of Currencies and Their Proxies}
    \label{fig7}
\end{figure}

Figure \ref{fig7} presents the mapping between proxies and BTC utilities. Token velocity serves as the proxy for the medium of exchange. It measures the percentage of tokens transacted over the past 24 hours relative to the total token supply. A large token velocity signifies frequent usage of and demand for the token.

\begin{figure}[h]
    \centering
    \includegraphics[width=\linewidth]{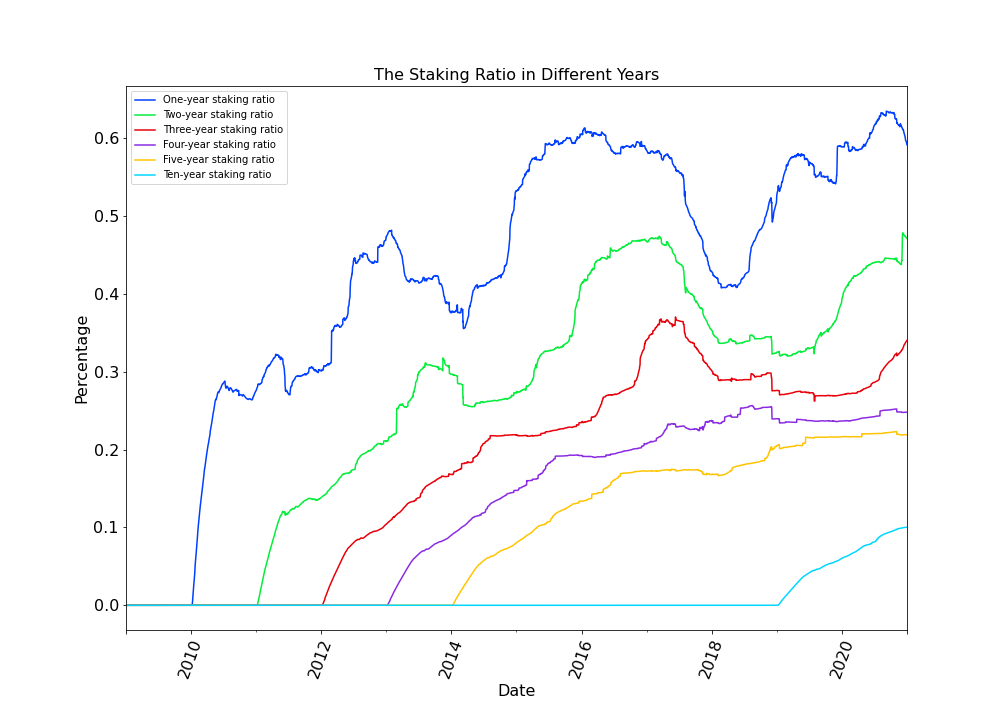}
    \caption{The Percentage of Staked BTC at 1, 2, 3, 4, 5, and 10 Years}
    \label{fig8}
\end{figure}

The Staking ratio is a proxy for the store of value, and it is defined as the percentage of tokens that are older than one year in age. As shown in Figure \ref{fig8}, a substantial amount of BTC tokens have been inactive for more than a year. These tokens serve as a store of value and can be referred to as “staked tokens.” A high staking ratio implies that more users are placing long-term faith in the Bitcoin system. Figure 8 shows the BTC staking ratio has been steadily increasing with periodic booms and busts.

In a nutshell, token velocity represents the utility of the token in its role as a medium of exchange that is demanded by daily active users, while the staking ratio represents the utility of the token as a store of value demanded by long-term holders.
The inverse of the dilution rate is another proxy for the store of value. The dilution rate measures the annual growth rate of the token supply. The new BTC comes from the block rewards. Ceteris paribus, a high dilution rate makes the token less attractive as a store of value, thus leading to a lower token utility.\footnote{The annualized dilution rate is defined as the 365 times moving average of the newly minted bitcoin in the last 90 days divided by the total bitcoin supply.} 

The inverse of price volatility\footnote{The 180/90/60/30-day volatility, measured as the standard deviation of the natural log of daily returns over the past 180/90/60/30 days.} is the proxy for the unit of account. Figure \ref{fig10} in the Appendix shows the volatility of BTC at 30, 60, 90, and 180 days. Note that due to the high volatility of the token price, most cryptocurrencies, except for stablecoins (see Senner and Sornette 2018) or coins with high stabilities \cite{gersbach2019},\footnote{Gersbach (2019)~\cite{gersbach2019} introduces flexible majority rules for the cryptocurrency issuance and shows that the flexible majority rules could foster the stability of a cryptocurrency.}  are limited in their utility as a unit of account. For simplicity, we will keep the price volatility mute in the following illustration of PU ratio; however, we include it in the robustness check in the machine learning session.

Figure \ref{fig11} in the Appendix illustrates the BTC token utility has been increasing in tandem with the BTC price. The growth of the BTC utility is mainly driven by the steady growth of the token staking ratio (see Figure~\ref{fig8}) and token velocity. The BTC price growth can be justified by the increase of its fundamentals (i.e., its token utility). Nevertheless, there are periods of deviations in the BTC price from its token utility, which typically indicates either overvaluation or undervaluation.

\begin{figure}[h]
    \centering
    \includegraphics[width=\linewidth]{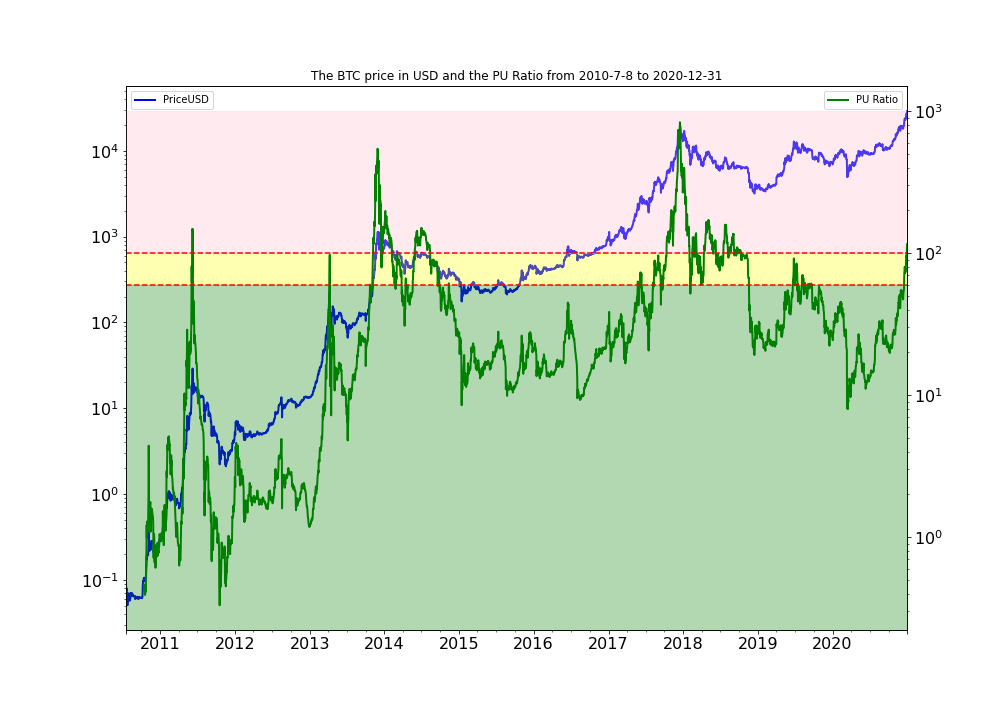}
    \caption{The BTC Price in USD (Blue Line, Left Axis) and PU ratio
(Green Line, Right Axis)}
    \label{fig12}
    \caption*{Note: Red zone (PU >100) indicates overvaluation, yellow zone (60 < PU < 100) indicates a normal range, and green zone (PU <60) indicates undervaluation.}
\end{figure}

Dividing BTC price by its utility yields the PU ratio of BTC (Figure~\ref{fig12}). Bitcoin’s first halving event in November 2012 broke the BTC supply-demand equilibrium and led to a price hike from around \$10 to \$100 in the first half of 2013. The BTC price soared from around \$100 in September 2013 to \$1,100 within a couple of months. Such a sharp price spike lured the long-term holders into selling their staked BTC (see the drop of BTC staking ratio in Figure \ref{fig8}). Correspondingly, the BTC token utility also plunged. The price hysteria, together with the significant drop in the token utility, drove the PU ratio to unsustainably high levels at the end of 2013, indicating the overvaluation of the BTC. Figure \ref{fig12} shows that the BTC price experienced a whole-year drop from above \$1,000 to around \$200 in 2014. From the beginning of 2015 to the mid-2016, the BTC price steadily increased from \$200 to around \$500 along with a firm growth of the BTC token utility (Figure \ref{fig11} in the Appendix), rendering a relatively stable PU ratio in the undervaluation range. The BTC price hike in this period can be justified by the increase in its fundamentals---a moderate increase in BTC velocity and staking ratio.
\par
After the second halving event in July 2016, the BTC dilution rate dropped to around 4\%. The formation of the new BTC supply-demand equilibrium led to the instability of the PU ratio between mid-2016 and mid-2017. The BTC price reached a historically high \$3,000 in June 2017. Long-term holders started to sell their BTC, leading to a drop in the BTC staking ratio in the second half of 2017. Meanwhile, the BTC velocity experienced a moderate drop. These two factors led to the drop in token utility. However, the BTC price rapidly grew to \$20,000. The drop in BTC token utility and skyrocketing BTC price resulted in a sharp spike in the Bitcoin PU ratio at the end of 2017. It took a whole year for the price to consolidate to around \$6,000 in 2018. The price even plunged to \$3,100 in early 2019, which was dubbed “crypto winter.” Since March 2019, an increase in token utility resulted from the rise of the token velocity and the staking ratio, along with the decrease in the dilution rate. Increasing demand boosted the BTC price to around \$15,000 in the summer of 2019, which signifies the end of the bearish crypto winter. The BTC price hovered around \$10,000 over the next year and achieved a new all-time high of \$30,000 at the end of 2020. The PU ratio once again poked into the yellow zone, which indicates a slight overvaluation of the BTC price.

\section{Cryptocurrency Valuation Ratio: Predictive Power on Future Returns}
In this section, we document the predictive regressions of BTC future 1-week, 30-day, 90-day, 180-day, and 360-day returns on proxies for BTC fundamental-to-market ratio. The proxies for the BTC valuation ratio include the inverse of the PE ratio (EPR), the inverse of the NVT ratio (TVN), the inverse of the P/M ratio (MPR), and the inverse of the PU ratio (UPR). To make a comparison with the results reported by Liu and Tsyvinski (2021), we also include the (negative) past 100$-$week cumulative BTC returns (Past100), the AMR,\footnote{The number of active addresses to market cap.} the TMR,\footnote{The number of transaction counts to market cap.}  PMR,\footnote{The number of payments (transfers) to market cap.}  and the first principal component of the previous eight proxies (FPC). 
We regress the BTC returns on the lagged cryptocurrency fundamental to market ratios, and the results are reported in Table \ref{table2}. Our regression differs from Liu and Tsyvinski (2021)~\cite{liu2021risks} in three aspects. First, we add four more fundamental-to-market ratios: EPR, TVN, MPR, and UPR. Second, we test the predictive power on long-horizon returns up to $360$ days while Liu and Tsyvinski (2021)~\cite{liu2021risks} only test for up to $8$ weeks. Third, our ratios are at daily frequencies while theirs are at weekly frequencies. We have new findings aside from those consistent with existing literature. First of all, none of these ratios has enough predictive power on future BTC returns in the short run (i.e., $1$-week and $30$-day return on investment (ROI)\footnote{In Table \ref{table2}, we can see that the R squared values in the first two columns are all below $6\%$.} ), which supports the finding in Liu and Tsyvinski (2021)~\cite{liu2021risks}. Second, AMR, PMR, TMP, and UPR together provide a good prediction of future BTC returns at a longer scale and the predictive power increases with longer horizons. Third, UPR best predicts long-term returns at $90$-day, $180$-day, and $360$-day horizons. The latter two new findings lead to a supplementary conclusion from Liu and Tsyvinski (2021)~\cite{liu2021risks}. We can conclude that there is a strong relationship between the future BTC returns and several fundamental-to-value ratios, especially the reverse of the PU ratio, at a longer horizon.
    
\begin{table}[!htbp]
    \centering
    \tiny
    \caption{Comparison of the Cryptocurrency Valuation Indicators}
    \label{table2}
    \begin{tabular}{cccccc}
    \\
    \hline
         & Future & Future & Future & Future & Future  \\
         & 1-week ROI & 30-days ROI & 90-days ROI & 180-days ROI & 360-days ROI \\
         \hline
         Past100 & -0.036 & -0.137** & 0.361*** & 0.663** & -2.217* \\
         & (-1.229) & (-1.663) & (3.144) & (2.822) & (-2.260)\\
         $R^2$ & 0.002 & 0.002 & 0.002 & 0.002 & 0.001 \\
         \hline
         AMR & 0.084*** & 0.448*** & 2.813*** & 7.521*** & 44.139*** \\
         & (4.229) & (7.095) & (5.982) & (9.370) & (10.711)\\
         $R^2$ & 0.007 & 0.017 & 0.076 & 0.197 & 0.347 \\
         \hline
         TMR & 0.253** & 1.000*** & 3.587*** & 7.285*** & 33.053*** \\
         & (6.295) & (8.371) & (8.289) & (12.205) & (10.173)\\
         $R^2$ & 0.038 & 0.052 & 0.075 & 0.112 & 0.118 \\
         \hline
         PMR & 0.102*** & 0.614*** & 3.767*** & 9.286*** & 56.428*** \\
         & (4.172) & (7.068) & (6.478) & (9.343) & (10.688)\\
         $R^2$ & 0.007 & 0.022 & 0.093 & 0.206 & 0.388 \\
         \hline
         EPR & 0.060* & 0.327*** & 1.174*** & 4.755** & 15.392*** \\
         & (3.523) & (5.162) & (6.632) & (9.794) & (12.045)\\
         $R^2$ & 0.007 & 0.018 & 0.026 & 0.154 & 0.082 \\
         \hline
         TVN & 0.032 & 0.144*** & 1.586*** & 6.925*** & 19.048*** \\
         & (1.318) & (2.160) & (8.783) & (14.654) & (10.754)\\
         $R^2$ & 0.002 & 0.003 & 0.038 & 0.260 & 0.101 \\
         \hline
         MPR & 0.016 & -0.021 & -0.202 & 0.184 & 5.906*** \\
         & (0.845) & (-0.321) & (-1.253) & (0.652) & (5.595)\\
         $R^2$ & 0.000 & 0.000 & 0.000 & 0.000 & 0.007 \\
         \hline
         UPR & 0.071*** & 0.455** & 3.906*** & 8.159*** & 49.486*** \\
         & (4.816) & (8.245) & (8.624) & (15.006) & (17.897)\\
         $R^2$ & 0.007 & 0.026 & 0.211 & 0.335 & 0.628 \\
         \hline
         FPC & 0.045*** & 0.238*** & 1.334*** & 3.873*** & 18.349*** \\
         & (5.111) & (8.166) & (7.158) & (11.990) & (11.974)\\
         $R^2$ & 0.010 & 0.025 & 0.086 & 0.264 & 0.302 \\
         \hline
    \end{tabular}
    \caption*{Notes: This table reports the predictive regressions of BTC future 1-day, 1-week, 30-day, 90-day, 180-day, and 360-day returns on proxies for BTC fundamental-to-market ratio. The proxies for BTC valuation ratio include the (negative) past 100-week cumulative BTC returns, the address-to-market ratio, the transaction-to-market ratio, the payment-to-market ratio, the inverse of PE ratio (EPR), the inverse of NVT ratio (TVN), the inverse of PM ratio (MPR), the inverse of PU ratio (UPR), and the first principal component (FPC) of the previous eight proxies. The Newey-West adjusted t-statistics with $n-1$ lags are reported in parentheses. *, **, and *** denote significance at the 10\%, 5\%, and 1\% levels. The data frequency is daily.}
\end{table}

\section{Cryptocurrency Valuation: Explainable Machine Learning}
An asset is overvalued (undervalued) when the ratio of the market cap to its fundamentals is high (low). The ratio is then useful in informing investors to follow the golden rule—“buy low and sell high.” In this section, we design K-means Clustering \cite{macqueen1967some}, an unsupervised machine learning method to assess further the explainability of the PU ratio in asset valuation. We identify a pattern: A high ROI is associated with a strategy of buying when the PU ratio is low and selling when the PU ratio is high, which is consistent with the theoretical implication of the PU ratio. 
\par
We represent each N-day investment by its PU ratios at buy-in and sell-out date:
\begin{equation}
    x^{(i)} = (PU Ratio_{buy-in} - PU Ratio_{sell-out} ) \label{2} 
\end{equation}
Then we label the data in four clusters. The indicator’s theoretical implication is deemed consistent with empirical data if the following two criteria are satisfied:

\begin{enumerate}
    \item 	There exists a cluster $k^*$ such that the PU ratio at the buy-in date is low and the PU ratio at the sell-out date is high; that is, for the centroid $\mu_{k^*}$ of cluster $k^*$, the map of $\mu_{k^*}$ on the x-axis is larger than that on the y-axis: $\mu_{k^*}|x<\mu_{k^*}|y$.
    \item	The investments in cluster $k^*$ have a higher ROI than investments in other clusters.
\end{enumerate}
\begin{figure}[!htbp]
    \centering
    \includegraphics[width=\linewidth]{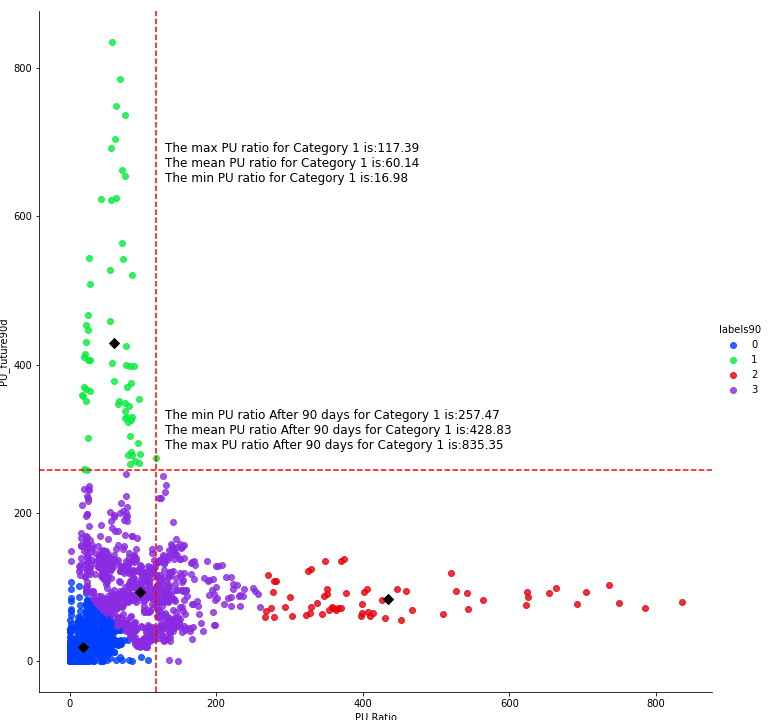}
    \caption{A: PU Ratio Clustering (90-Day ROI)}
    \includegraphics[width=\linewidth]{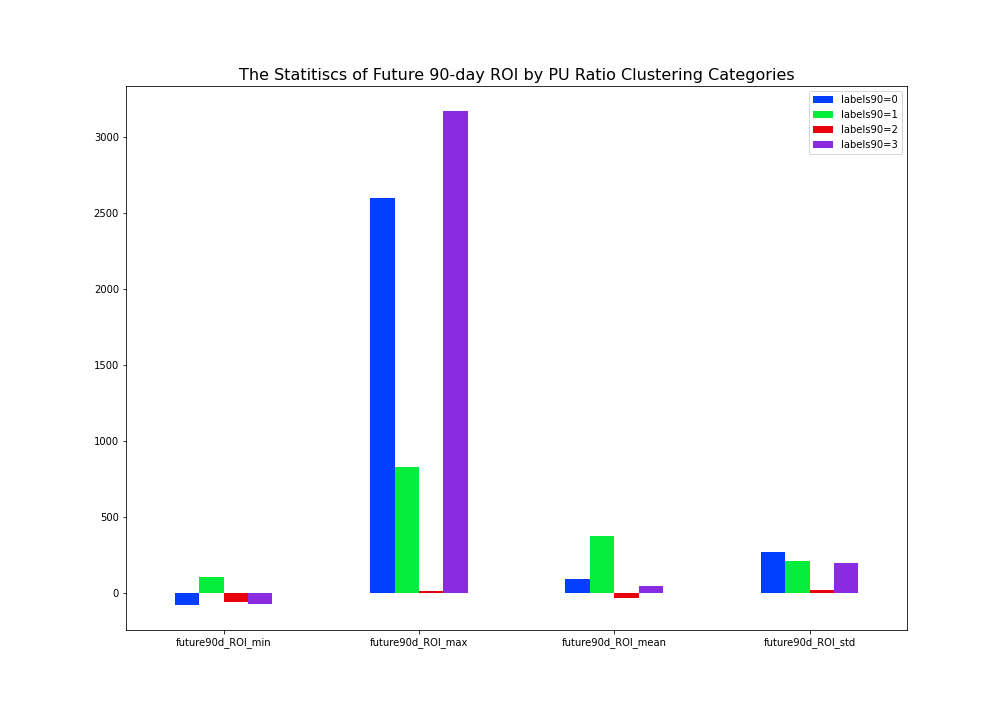}
    \caption*{B: 90-Day ROI by Clustering Categories (PU Ratio)}
    \label{fig13}
\end{figure}

Figure \ref{fig13} shows that only the PU ratio satisfies the two criteria. In Figure \ref{fig13} A, the x-axis represents the PU ratio on the buy-in date and the y-axis represents the PU ratio on the sell-out date. Figure \ref{fig13} B shows the ROI by labeled clusters. Cluster 1 in green manifests a pattern of buy-low (highest PU ratio on buy date: 117.39) and sell-high (lowest PU ratio after 90 days: 257.47) and the mean of its ROI dominates the three other clusters. Our results are robust in varying trading periods. We fail to identify similar patterns for the NVT and PM ratios. 

\section{Automated Trading Strategies}    
In this section, we propose an automated trading strategy based on the PU ratio and compare its performance to a conventional buy-and-hold strategy and a moving average (MA) crossover rule in finance\footnote{The moving average rules (Gartley 1935~\cite{gartley1935profits}) give a buy (sell) signal when the short-window moving average of the current price indicator moves above (below) its long-window moving average. Literature in finance (e.g., Brown and Jennings 1989~\cite{brown1989}; Brock, Lakonishok, and Lebaron 1992~\cite{brock1992simple}, Neely et al. 2014~\cite{neely2014forecasting}) show evidence that investors benefit more by following the moving average rules than the buy-and-hold strategies. The intuition is that the moving average rules predict a short-term positive fluctuation in price when its short-term indicator moves above (below) its long-term indicator.}.  We find that the strategy based on the PU ratio outperforms the other two conventional strategies.
\par
\begin{figure}[!htbp]
    \centering
    \includegraphics[width=\linewidth]{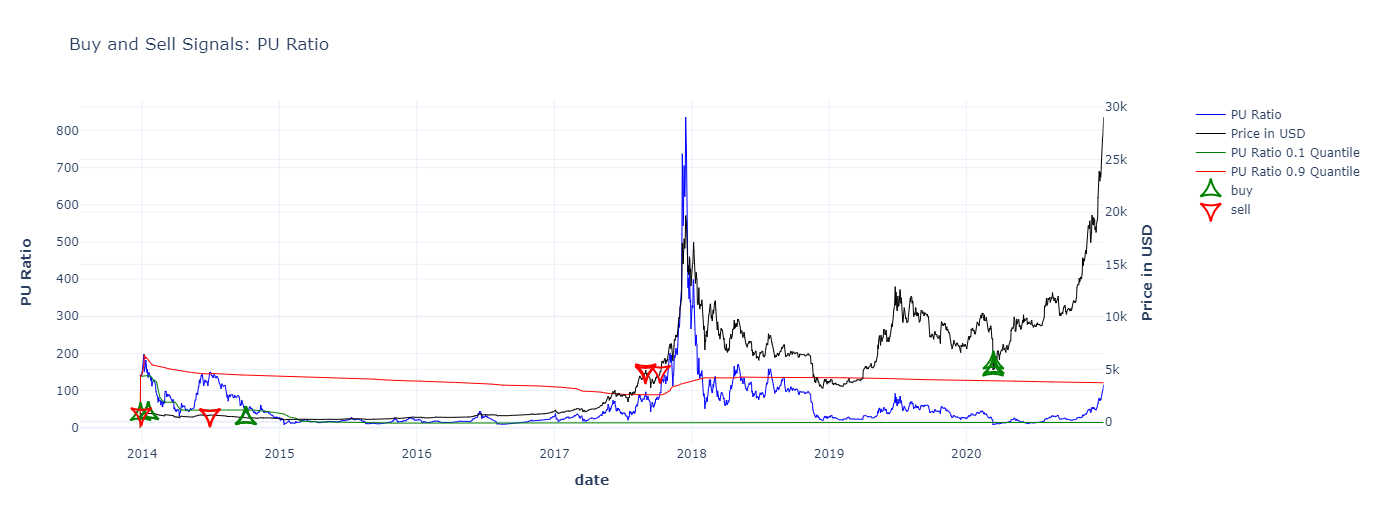}
    \caption{A: Buy and Sell Signals for the PU Ratio}
    \includegraphics[width=\linewidth]{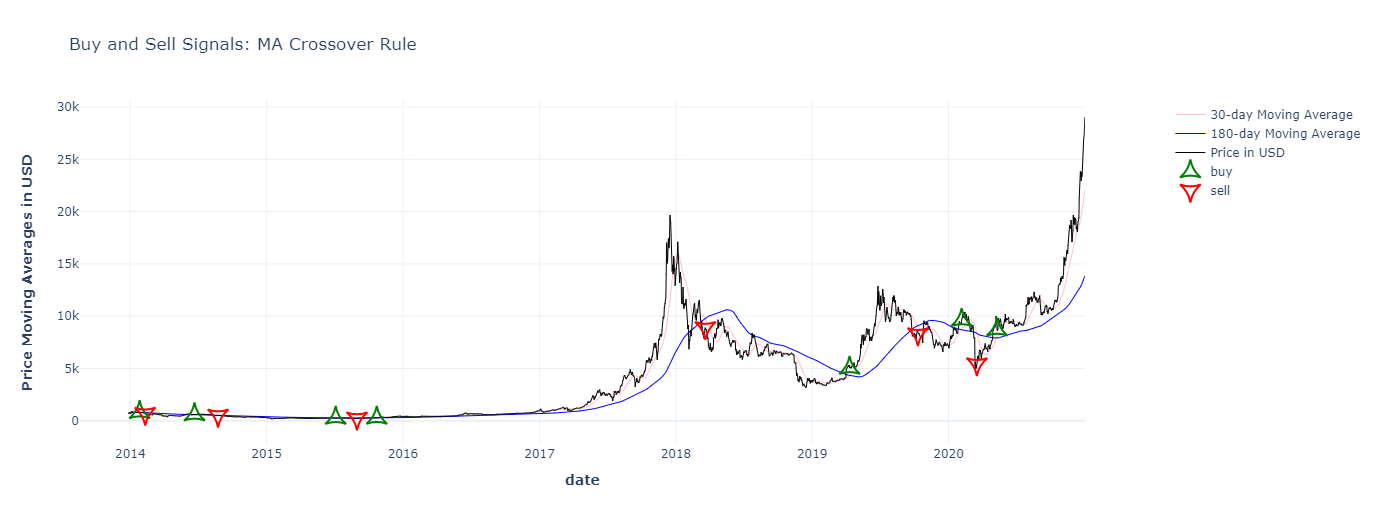}
    \caption*{B: Buy and Sell Signals for the MA Crossover Rule}
    \includegraphics[width=\linewidth]{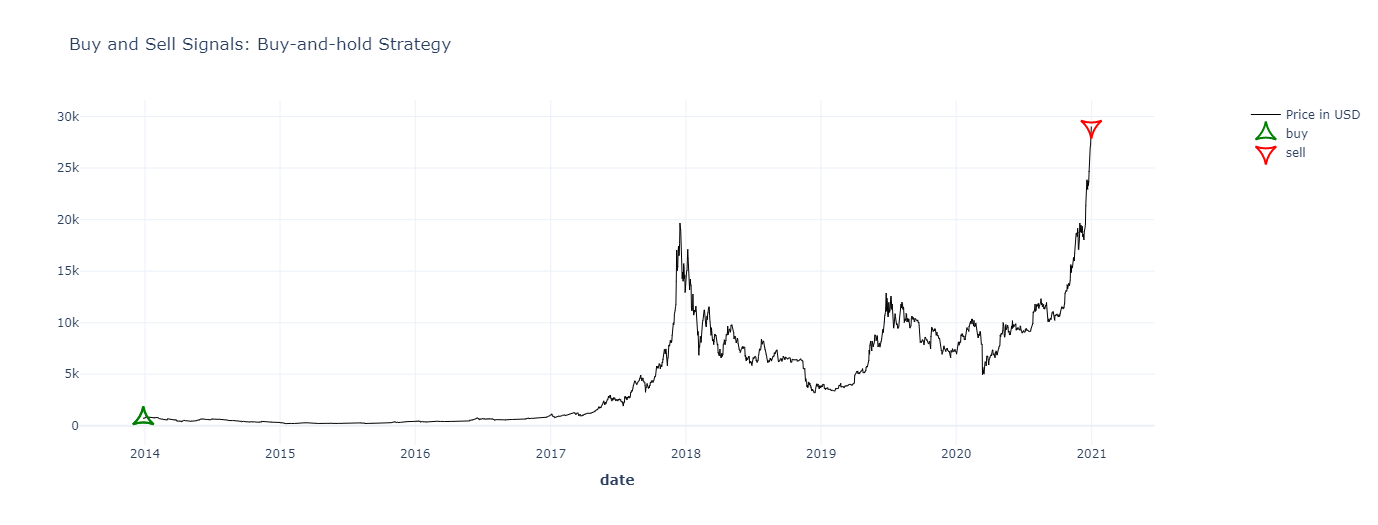}
    \caption*{C: Buy and Sell Signals for the Buy-and-Hold Strategy}
    \label{fig_20}
\end{figure}

Our strategy generates a buy (sell) signal when the PU ratio is equal to or lower (higher) than its 0.1 (0.9) quantiles of the historical records. The MA crossover rule generates a buy (sell) signal when the short-window MA of BTC price crosses up (down) its long-window MA. Finally, the buy-and-hold strategy buys at the start date and holds until the end date of the investment period. Figure \ref{fig_20} demonstrates the trading strategies with buy and sell signals. 
\par
We evaluate the performance of the three strategies with an initial capital of 100,000 USD, a 0.1\% transaction fee, and a transaction limit of 100 BTC at each buy or sell signal.\footnote{The 0.1\% transaction fee is aligned with the standard fee structure at Binance, the largest crypto exchange platform.}  We consider the timeframe from the first date of the BTC off-chain transaction on December 27, 2013, to December 31, 2020.  At the end of the trading day, the automated trading strategy based on the PU ratio generates a gross ROI of 6,245.83\% and an annualized Sharpe ratio of 3.65, which is higher than both the market timing strategy based on the MA crossover rule (gross ROI: 2,670.37\%; annualized Sharpe ratio: 3.29) and the buy-and-hold strategy (gross ROI: 3,920.09\%; annualized Sharpe ratio: 2.87). The result is robust in a variety of parameter settings and even better when we do not set a transaction limit. 

\section{Conclusion and Discussion}
Our results imply that considering features unique to the crypto market could contribute significantly to cryptocurrency valuation. For instance, in this study, we take advantage of Blockchain UTXO accounting methods to construct a proxy for store-of-value. This approach could be seamlessly applied to other UTXO blockchains such as Bitcoin Cash,\footnote{\href{https://www.bitcoincash.org/whitepaper}{https://www.bitcoincash.org/whitepaper}}  Dash,\footnote{\href{https://cryptoverze.com/dogecoin-whitepaper}{https://cryptoverze.com/dogecoin-whitepaper}}  Dogecoin,\footnote{\href{https://docs.dash.org/en/stable/introduction/about.html}{https://docs.dash.org/en/stable/introduction/about.html}}  Litecoin,\footnote{\href{https://www.allcryptowhitepapers.com/litecoin-whitepaper}{https://www.allcryptowhitepapers.com/litecoin-whitepaper}}  and Zcash.\footnote{https://whitepaper.io/coin/zcash}  Moreover, our approach can adapt to the updates of blockchain mechanisms and generate further research findings. In recent years, cryptocurrencies based on proof-of-work have been criticized as a waste of energy. Most of the electricity consumed by miners is converted into hashing power to compute cryptographic puzzles instead of routing and executing transactions. To address this issue, proof-of-stake (PoS) blockchain projects, such as EOS,\footnote{\href{https://www.allcryptowhitepapers.com/eos-whitepaper}{https://www.allcryptowhitepapers.com/eos-whitepaper}}  Tezos,\footnote{\href{https://wiki.tezosagora.org/whitepaper}{https://wiki.tezosagora.org/whitepaper}}  and Cosmos,\footnote{\href{https://v1.cosmos.network/resources/whitepaper}{https://v1.cosmos.network/resources/whitepaper}}  have gained momentum in recent years. In the PoS protocol, a certain fraction of native tokens are staked in the system by validators (i.e., miners, see Saleh 2020~\cite{saleh2021blockchain}). Validators do not need to solve cryptographic puzzles to win the block rewards. The more tokens staked by a validator, the higher the chance that the validator becomes a block maker and earns the block rewards. Moreover, PoS projects have a flexible token dilution rate. For example, the dilution rate is dynamically adjusted by the Cosmos system to achieve a 66.7\% staking ratio. When the staking ratio drops below the target, the block rewards automatically increase to attract more tokens staked in the system and vice versa. In such a case, the staking ratio and the dilution rate as proxies for store-of-value would be interdependent. Besides store-of-value, other features unique to the crypto market such as the decentralization level~\cite{zhang2022sok}, the network features of the peer-to-peer transactions~\cite{ao2022decentralized}, sentiments on blockchain security~\cite{fu2022ai} and the upgrade of blockchain mechanisms~\cite{eip1559,zhang2023understand} might also affect the values of cryptocurrency. It would be interesting to explore further in this direction.  

Like all other asset prices, cryptocurrency prices are also affected by macroeconomic development, which is beyond the scope of this paper. For example, on March 12, 2020, the cryptocurrency market crashed with the stock market and gold market due to the panic sale caused by the COVID-19 pandemic. Bitcoin prices plunged by more than 60\%, almost twice as much as the stock market (e.g., Dow Jones Industrial Average and S\&P 500 Index). However, one and half months after the market collapse, the BTC price quickly recovered to its pre-crisis level while the stock market still stagnated at around 80\% of its pre-crisis level. It would be interesting to study what unique attributes of the cryptocurrency market led to the different market trajectories and what the implications are for investment strategies.

In the current paper, the indicators that we develop for cryptocurrency valuation are based on macro variables and functions of currency in monetary theory. However, every macro phenomenon is a manifestation of micro behavior in an aggregate form, which in general is derived from two micro facets: the choices of each individual based respectively on rationality or heuristic bias contingent on the state of the world. For the first, to model rational choices, each individual must envision at least all hypothetical scenarios in Decision under Ambiguity \cite{karni2015ambiguity} and also form subjective probabilities for all possible scenarios in Decision under Uncertainty \cite{gilboa2009theory}. However, since crypto-economics is rapidly evolving, both requirements are difficult if not far-fetched for the majority, which makes the problems more similar to decisions under ignorance \cite{giang2016decision,karni2013reverse,hogarth1995decision,maskin1979decision}, a vastly uncultivated field even at the foundations of microeconomic theory. Second, little literature seeks to understand behavioral patterns of bounded rationality~\cite{conlisk1996bounded,levin2022bridging}. For instance, Gemayel and Preda (2021)~\cite{gemayel2021performance} find evidence of heuristic bias including disposition effect \cite{shefrin1985disposition}, self-attribution bias \cite{hoffmann2014self}, and the gambler’s fallacy \cite{chen2016decision} among cryptocurrency traders. We envision the two directions as representing promising directions and the next frontiers for future research.

Hey (2009)~\cite{hey2009fourth} envisions the fourth paradigm shift in scientific discoveries to be data-driven. Our study shows that interdisciplinary research in machine learning could verify the explainability of valuation ratios by establishing consistency for its theoretical implications in empirical results. To facilitate this future research, we distribute the trading algorithms as open-source software via Python Package Index \cite{adadi2018peeking} (PyPI).\footnote{refer to the latest version: \href{https://test.pypi.org/project/AlgorithmicTradingCV}{https://test.pypi.org/project/AlgorithmicTradingCV}, and a related data science pipeline paper~\cite{Zhang2022IEEE}.}

\bibliographystyle{spmpsci}
\bibliography{CV}

\appendix
\newpage
\section{K-means Clustering Algorithms}
K-means clustering aims to partition n observations to k clusters in which each observation belongs to the cluster with the nearest cluster centroid. Specifically, given a data set $ {x^{(1)},x^{(2)},...,x^{(m)}}$, our goal is to predict $k$ centroids and a label $c^{(i)}$ for each data point, such that each cluster centroid’s $\mu_k$ is the closest to all data points with label $c^{(i)}=k$. 
The algorithm finally finds the best centroids and labels by alternating between (1) labeling data points in clusters based on the current centroids and (2) choosing centroids based on current cluster labels.
\begin{algorithm}[ht]
    \SetAlgoNoLine
    Initialize cluster centroids $\mu_1$,$\mu_2$,...,$\mu_k \in R^n$ randomly\;
    \While{not converge}{
    for every $i$, set $c^{(i)} = argmin_j||x^{(i)}-\mu_j||^2$\;
    for every $j$, set $\mu_j=\frac{\sum_{i=1}^m I[c^{(i)}=j]x^{(i)}}{\sum_{i=1}^m I[c^{(i)}=j]}$.}  
    \caption{K-means Clustering}
    \label{alg1}
\end{algorithm}

\section{Supervised Machine Learning}
In this section, we use Decision Tree Classifier \cite{breiman1984classification}, a supervised machine learning model with high explainability \cite{gunning2019xai} to explore the explainability of the PU ratio further. A PU ratio below an identifiable threshold effectively predicts a bull market. The finding is consistent with the theoretical implication that a low PU ratio implies an undervaluation, which predicts future market rebounds.
\subsection{Decision Tree Classifier}
Decision Tree Classifier aims to recursively split the dataset into targeting labeled classes according to relevant features such that the observed labels are as pure as possible (majority to be only 1 or 0) for the resulting sub-datasets after splits. Each split is represented by a node and the splitting criteria are denoted along a branch of the tree.
\begin{algorithm}[t]
    \SetAlgoNoLine
    1. Pick a criterion to evaluate information gain, e.g. $entropy$ or $gini$\;
    2. Pick targeting labels and relevant features\;
    \While{not converge}{
    a. find a descriptive feature such that the optimal split along this feature achieves the highest information gain\;
    b. split the dataset along the descriptive feature using the optimal split.}  
    \caption{Decision Tree Classifier}
    \label{alg2}
\end{algorithm}
Entropy and Gini are two criteria for calculating information gains. A sub-dataset with multiple observed labels is impure, whereas the one with only one label is pure. Let $p(c_i)$denote the percentage/probability of observed label $c_i$ in a (sub) dataset, and the formulas for Entropy and Gini are:
\begin{equation}
    Entropy = \sum_{i=1}^n -\ p(c_i)\ log_2\ (p(c_i)) \label{3} 
\end{equation}
\begin{equation}
    Gini = 1 - \sum_{i=1}^n p^2 (c_i) \label{4} 
\end{equation}
The information gained for each split is the difference in Entropy (Gini) before and after the split. The Entropy (Gini) after each split is the weighted Entropies (Ginis) of resulting sub-datasets, where weights are the percentage of observations.

\subsection{Method and Result}
\begin{figure}[!htbp]
    \centering
    \includegraphics[width=0.5\linewidth]{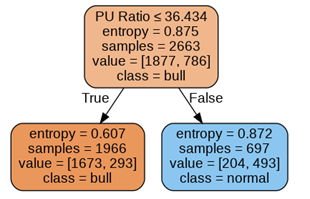}
    \caption{Bull Market Decision Tree Prediction for 180-Day Investments (Criterion=Entropy, Maximum Depth = 1)}
    \caption*{Note: Each note in the figure includes: (1) the criteria for splitting; (2) entropy for the sub-datasets; (3) sample size; (4) the sample size for the bull market or not; and (5) the class label. For example, in the root note, the criterion for splitting is whether the PU ratio at the buy-in date is smaller than 36.434. The training dataset has 2663 observations with 1,877 bull market labels.}
    \label{fig19}
\end{figure}

To further test the consistency of the PU ratio, we predict the bull market with the PU ratio at the buy-in date: $I[bull \ market] \sim f(PU \ ratio)$. We adopt the most common definition of a bull market: when the ROI for investment is larger than $20\%$\footnote{ \href{https://www.investopedia.com/terms/b/bullmarket.asp}{https://www.investopedia.com/terms/b/bullmarket.asp}}.  We split the earlier $75\%$ data to train the model and use the rest for testing. \footnote{The results are robust using entropy instead of Gini across different random states, and the current random state is zero. We restrict the maximum depth for illustration simplicity.}.  The result in Figure \ref{fig19} shows a significant bull market prediction when the PU ratio is smaller than 36.434. The results are consistent with the economic intuition: When the PU ratio is low, the asset is undervalued and thus promising to generate high returns. More specifically, the result suggests a cutoff strategy of buying when the PU ratio is smaller than $36.434$, which will very likely generate a return of over $20\%$ in $180$ days. The result is robust when we change the criterion for the quality of splits to Gini and the duration of investments. However, the precision of the prediction decreases for shorter periods of investment, indicating that the PU ratio is an indicator that serves long-term investments better.

\section{Additional Tables and Figures}
\label{atf}
\begin{figure}[!htbp]
    \centering
    \includegraphics[width=\linewidth]{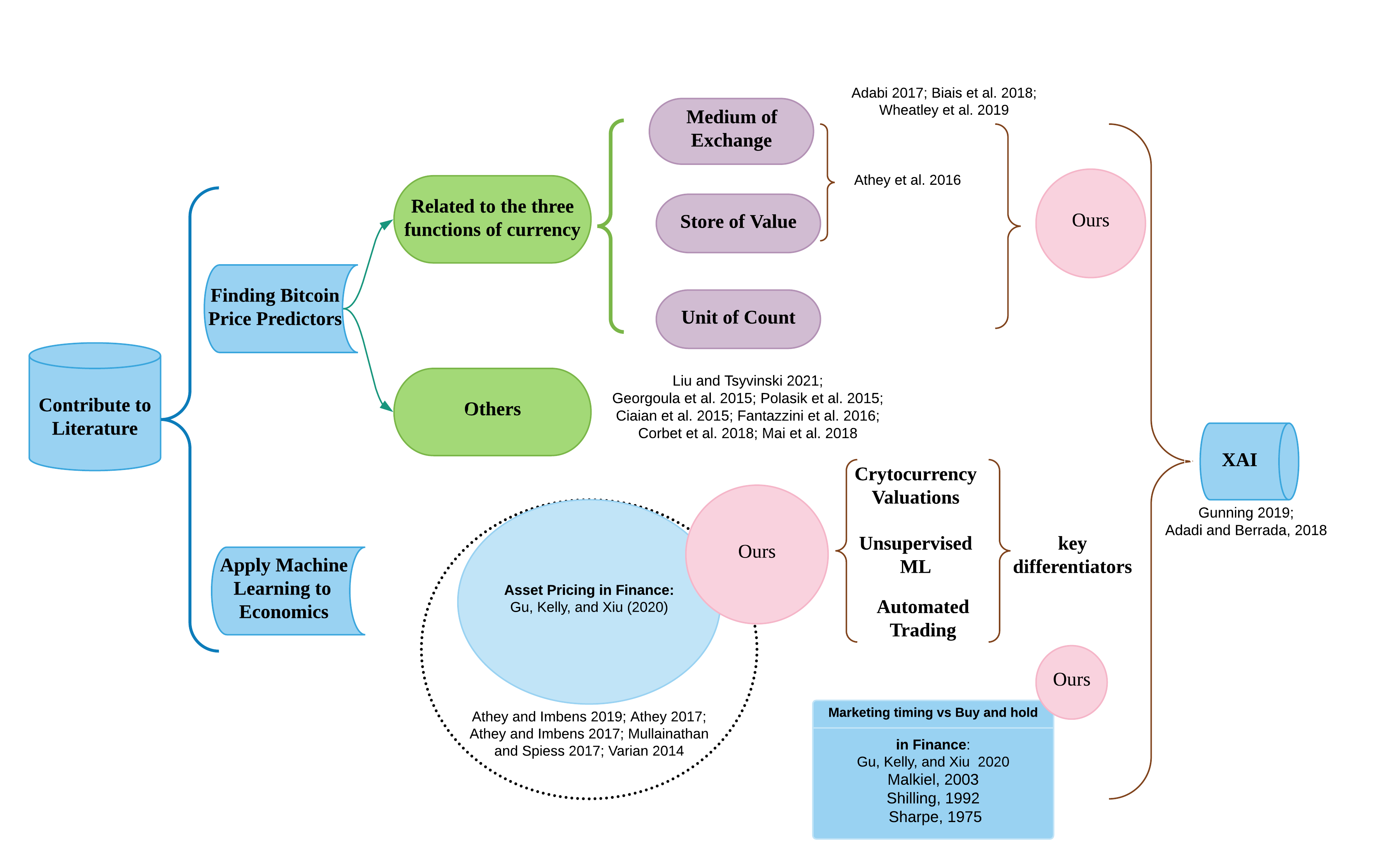}
    \caption{Contribution to the Existing Literature}
    \label{fig1}
\end{figure}

\begin{figure}[!htbp]
    \centering
    \includegraphics[width=\linewidth]{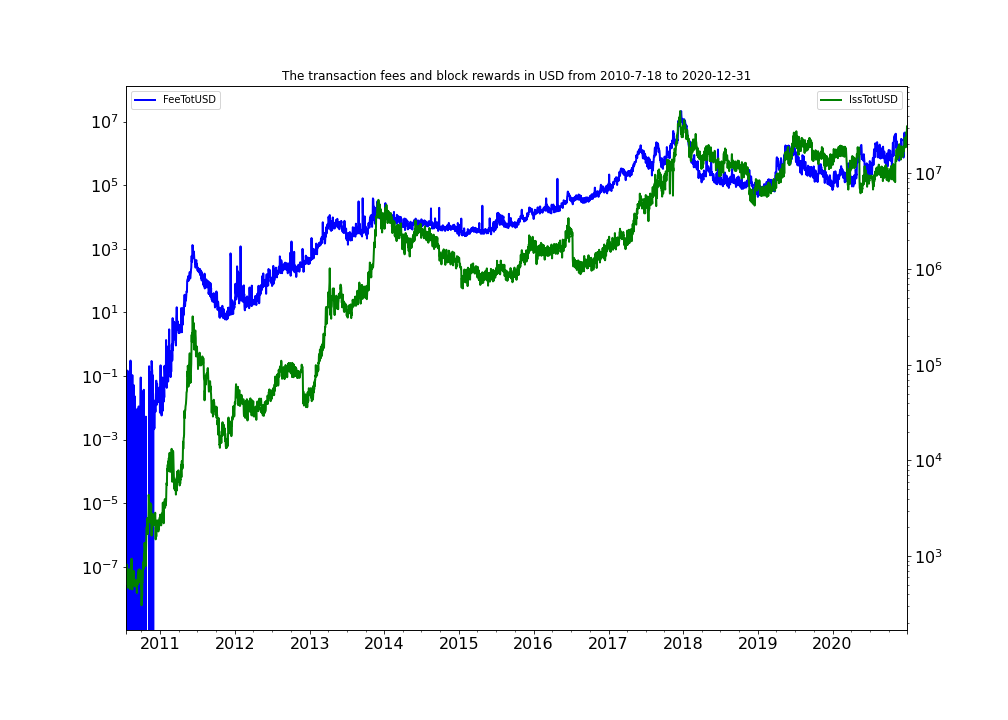}
    \caption{Miners’ Revenue in USD: Transaction Fees and Block Rewards}
    \label{fig3}
\end{figure}

\begin{figure}[!htbp]
    \centering
    \includegraphics[width=\linewidth]{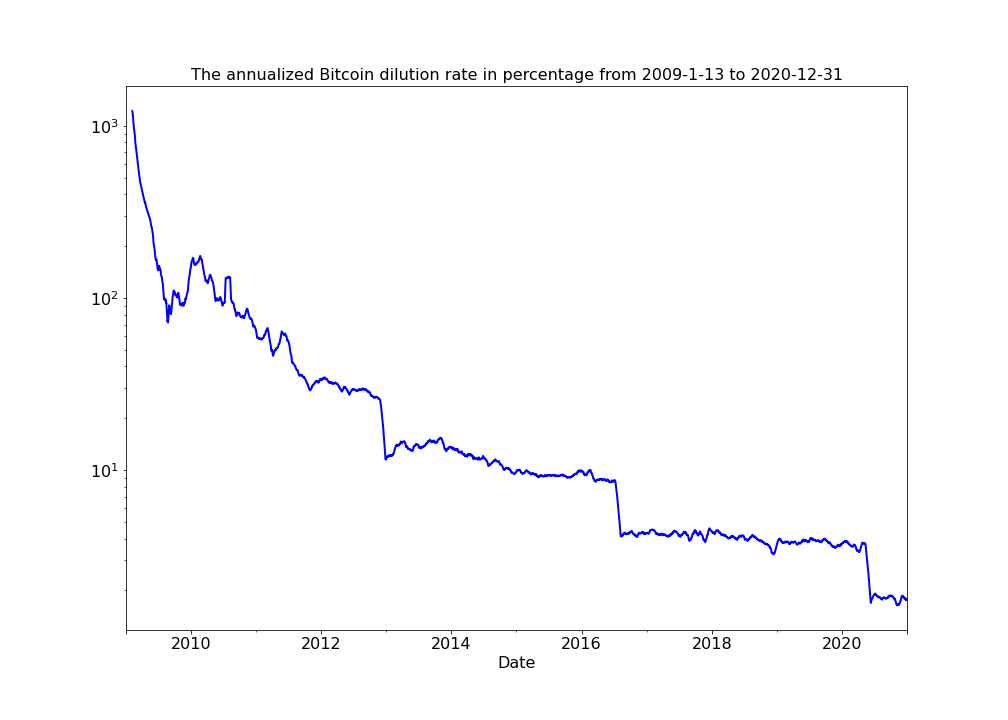}
    \caption{Annualized BTC Dilution Rate}
    \label{fig9}
\end{figure}

\begin{figure}[!htbp]
    \centering
    \includegraphics[width=\linewidth]{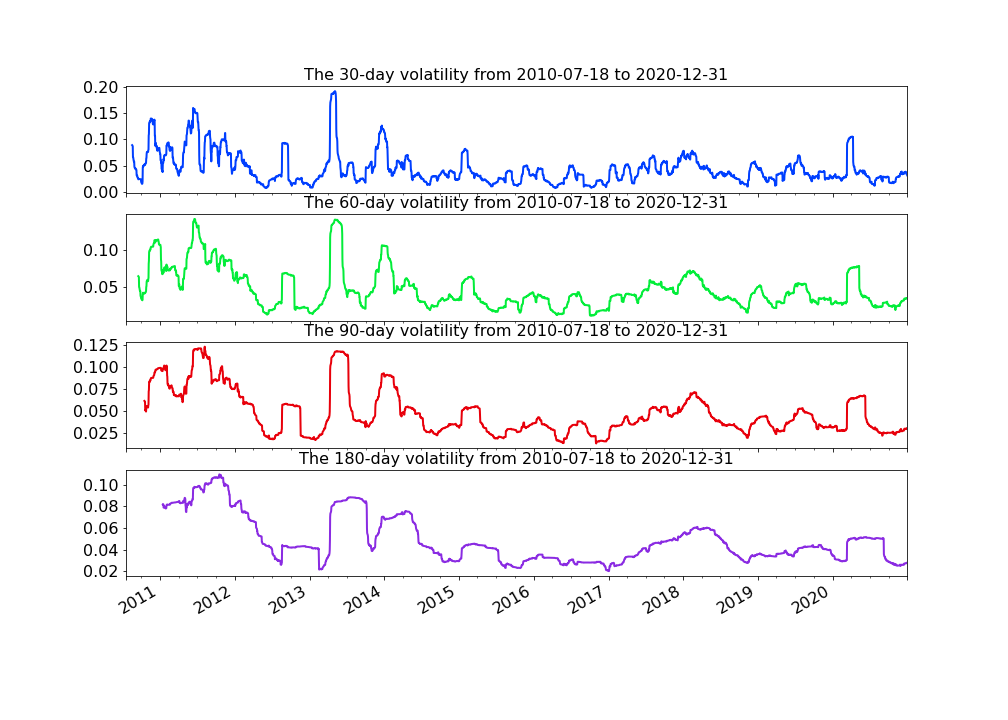}
    \caption{The BTC Volatility at 30, 60, 90, and 180 Days}
    \label{fig10}
\end{figure}

\begin{figure}[!htbp]
    \centering
    \includegraphics[width=\linewidth]{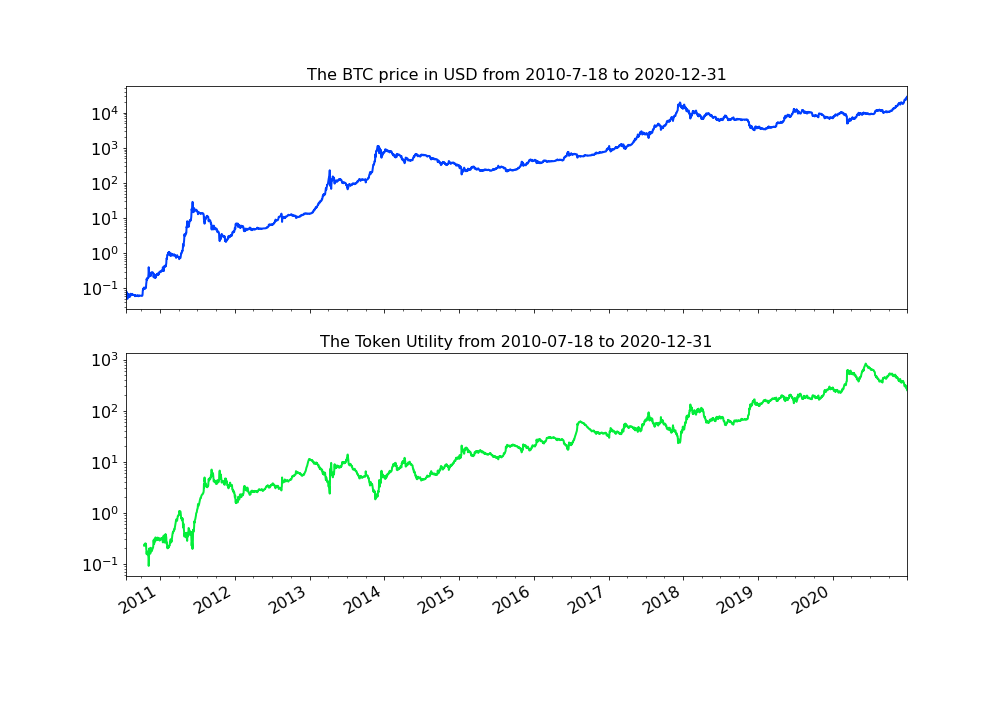}
    \caption{The BTC Price in USD (Blue Line) and Token Utility (Green Line)}
    \label{fig11}
\end{figure}
    
\begin{figure}[!htbp]
    \centering
    \includegraphics[width=\linewidth]{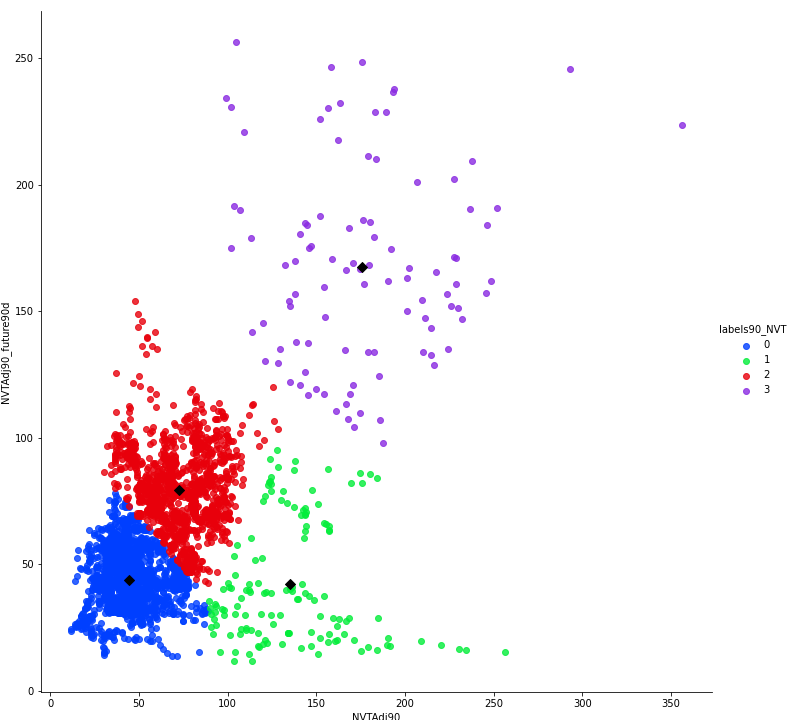}
    \caption{A: NVT Ratio Clustering (90-Day ROI)}
    \includegraphics[width=\linewidth]{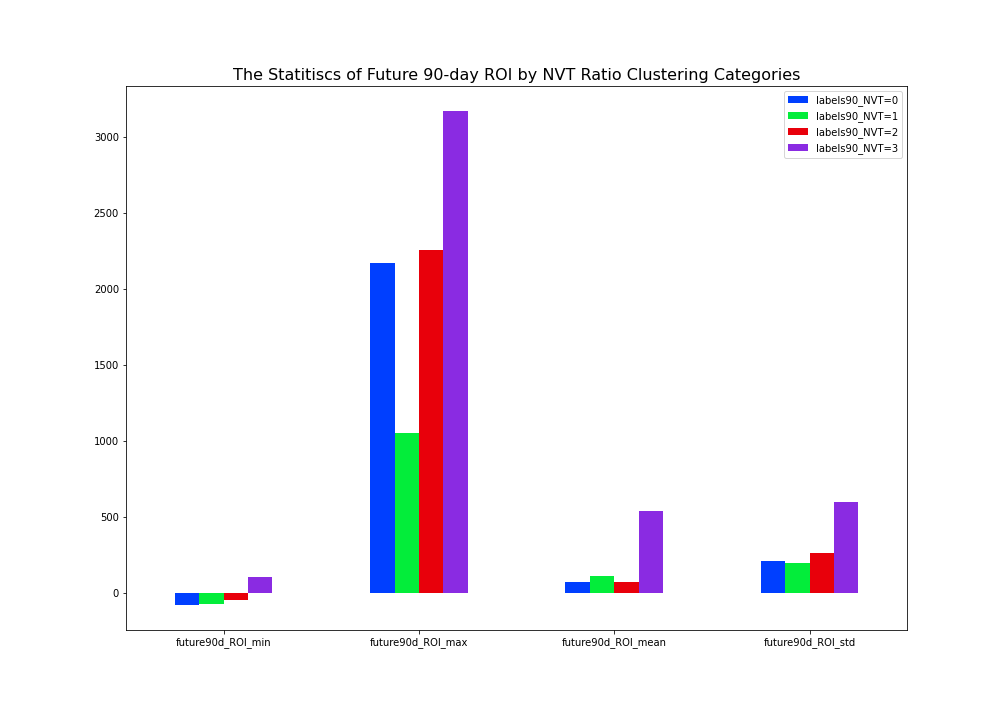}
    \caption*{B: 90-Day ROI by Clustering Categories (NVT Ratio)} 
    \label{fig15}
\end{figure}

\begin{figure}[!htbp]
    \centering
    \includegraphics[width=\linewidth]{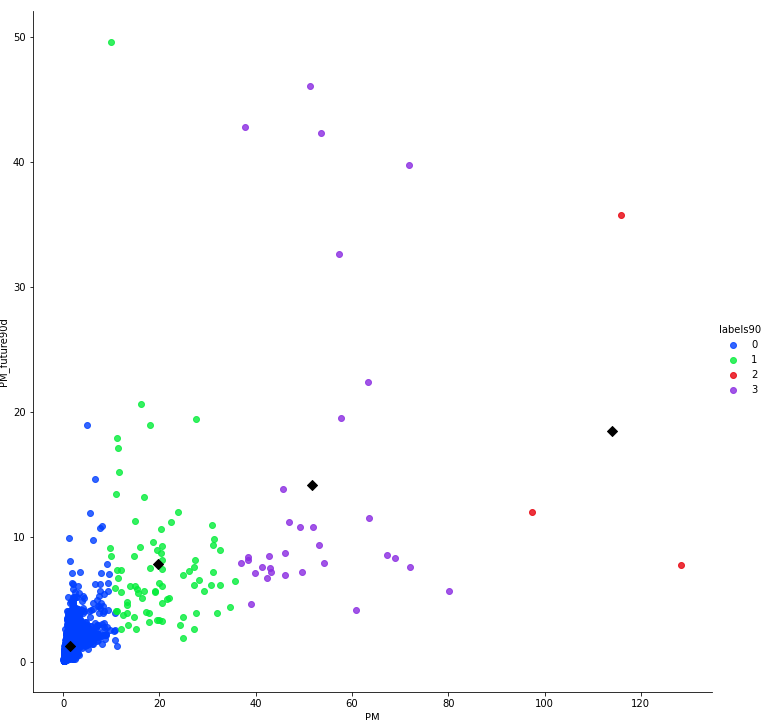}
    \caption{A: PM Ratio Clustering (90-Day ROI)}
    \includegraphics[width=\linewidth]{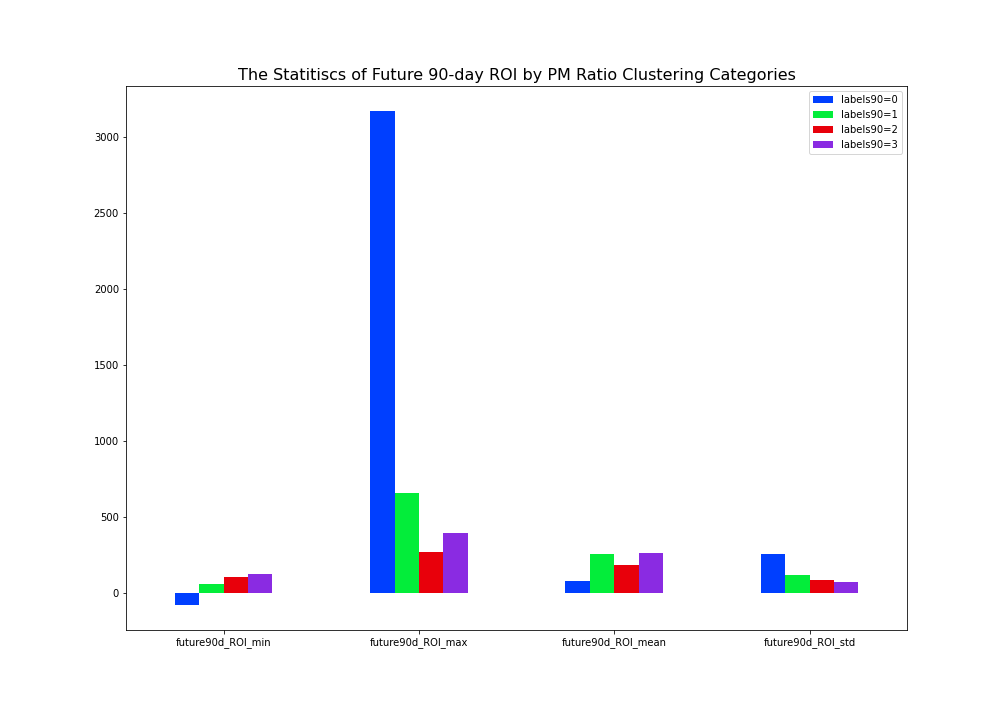}
    \caption*{B: 90-Day ROI by Clustering Categories (PM Ratio)}
    \label{fig17}
    \end{figure}

\begin{figure}[!htbp]
    \centering
    \includegraphics[width=\linewidth]{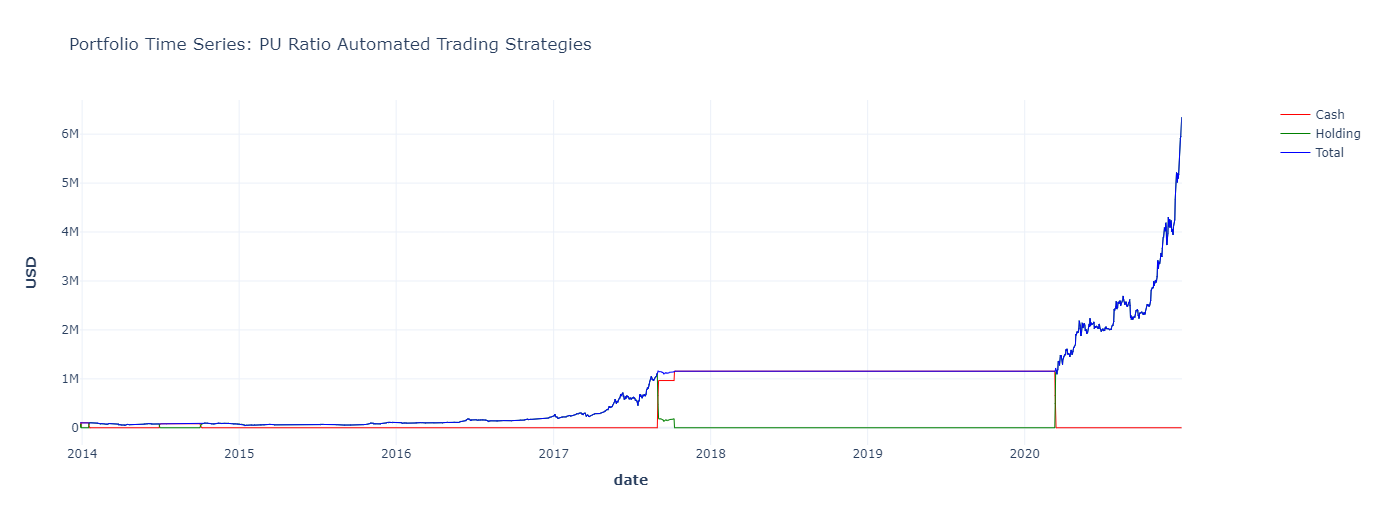}
    \caption{Portfolio Time Series for the PU Ratio Automated Trading Strategies (Gross ROI: 6,245.83\%, Annualized Sharpe Ratio: 3.65)}
    \label{fig_23}
\end{figure}  

\begin{figure}[!htbp]
    \centering
    \includegraphics[width=\linewidth]{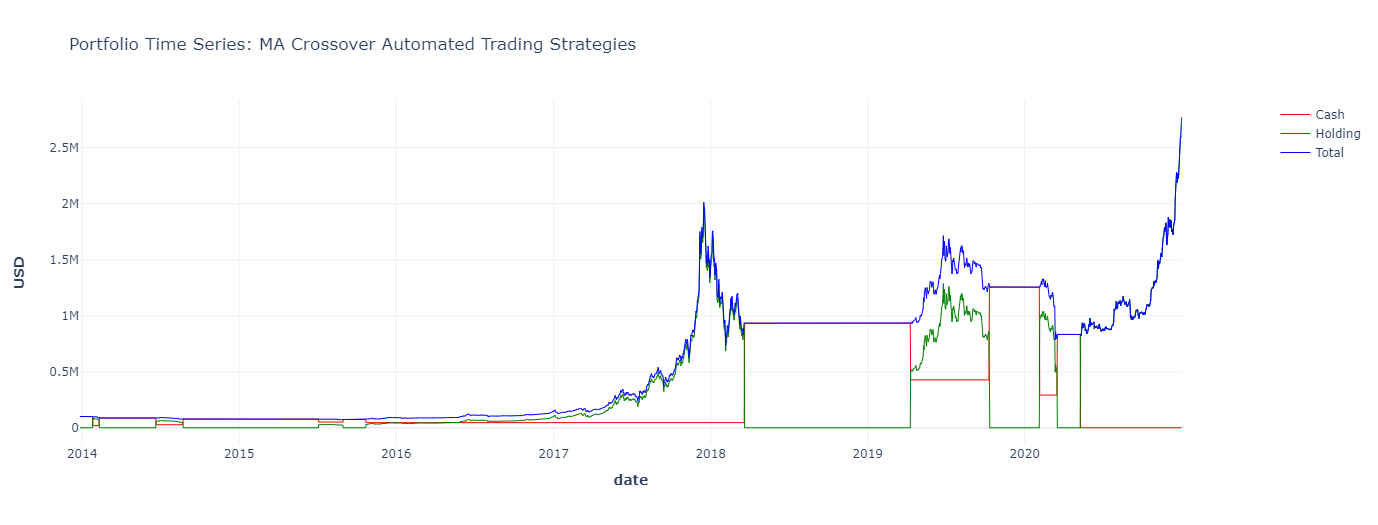}
    \caption{Portfolio Time Series for the MA Crossover Rule
(Gross ROI: 2,670.37\%; Annualized Sharpe Ratio: 3.29)}
    \label{fig_24}
\end{figure} 
    
\begin{figure}[!htbp]
    \centering
    \includegraphics[width=\linewidth]{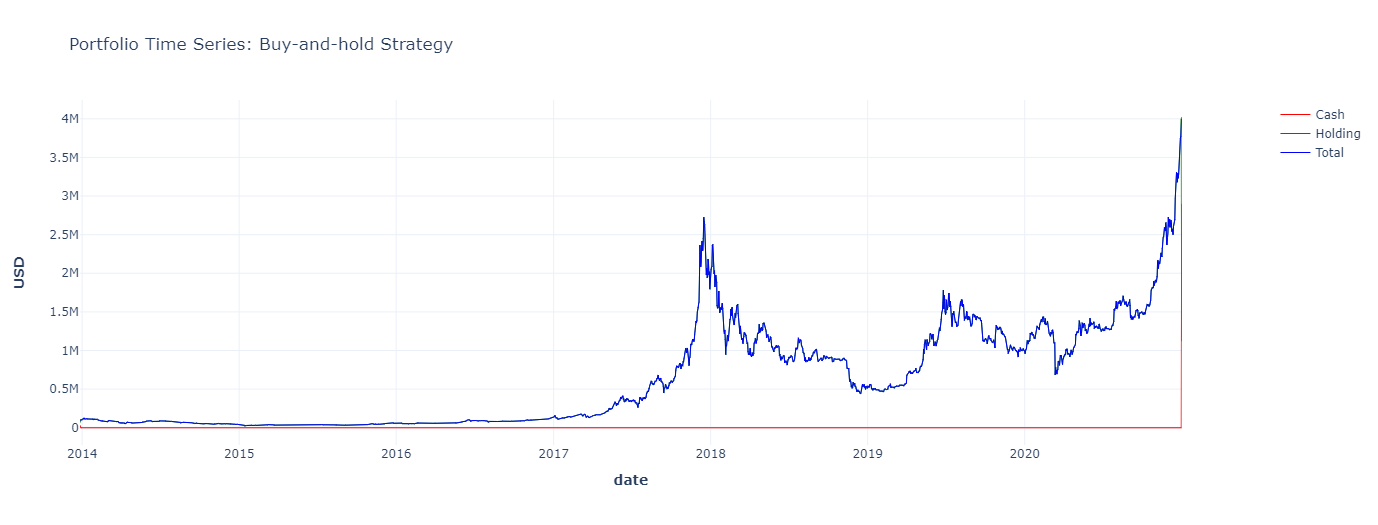}
    \caption{Returns for the Buy-and-Hold Strategy (ROI: 2,899.36\%)}
    \label{fig_25}
\end{figure}  
    
\begin{figure}[!htbp]
    \centering
    \includegraphics[width=\linewidth]{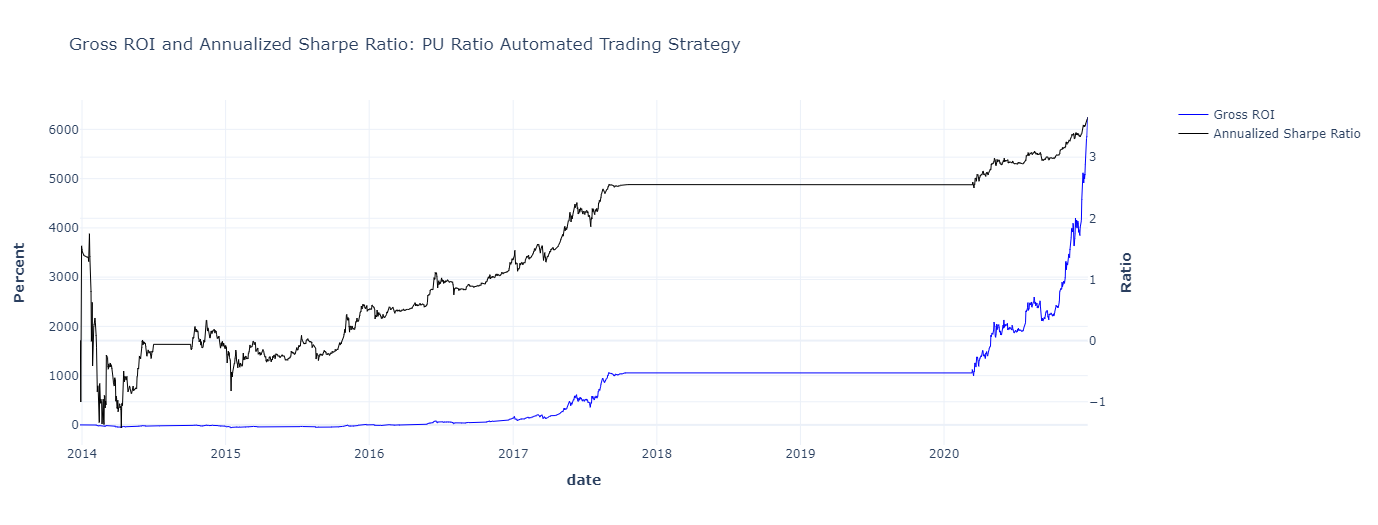}
    \caption{Gross ROI and Annualized Sharpe Ratio Time Series for the PU Ratio}
    \label{fig26}
\end{figure} 
    
\begin{figure}[!htbp]
    \centering
    \includegraphics[width=\linewidth]{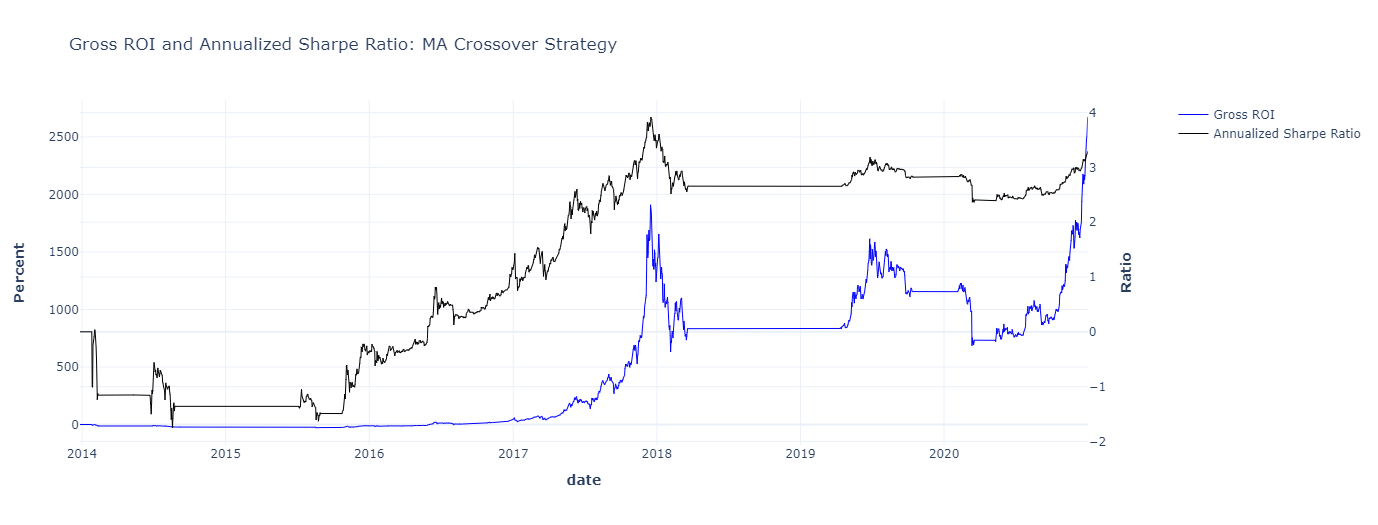}
    \caption{Gross ROI and Annualized Sharpe Ratio Time Series for the MA Crossover Rule}
    \label{fig27}
\end{figure} 
    
\begin{figure}[!htbp]
    \centering
    \includegraphics[width=\linewidth]{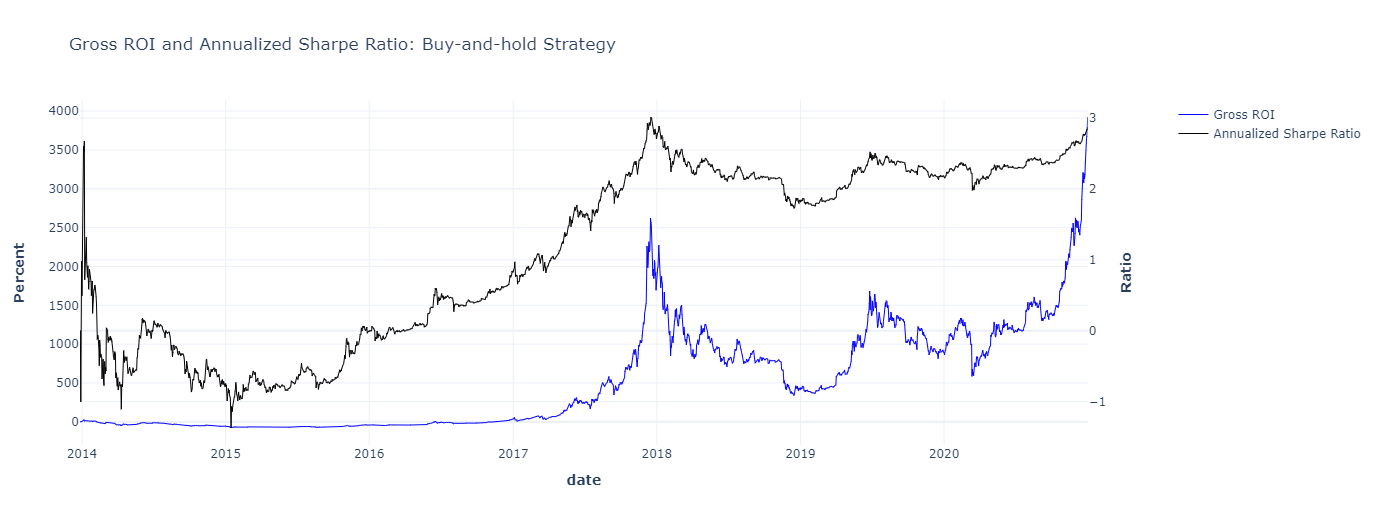}
    \caption{Gross ROI and Annualized Sharpe Ratio Time Series for the Buy-and-Hold Strategy}
    \label{fig28}
\end{figure} 

\begin{figure}[!htbp]
    \centering
    \caption{Correlation of fundamental-to-market ratio (at daily frequency)}
    \includegraphics[width=\linewidth]{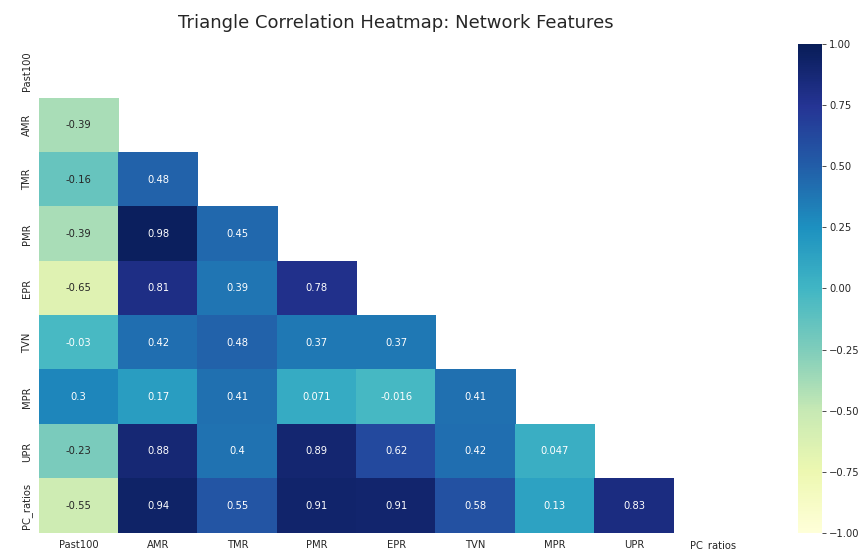}
    \label{heatmap}
    \end{figure}

\end{document}